# CNN-based Surface Temperature Forecasts with Ensemble Numerical Weather Prediction

Takuya Inoue, Takuya Kawabata

*Meteorological Research Institute, Tsukuba, Ibaraki, Japan*

*Corresponding author*: Takuya Inoue, inoue@mri-jma.go.jp

ABSTRACT

Due to limited computational resources, medium-range temperature forecasts typically rely on low-resolution numerical weather prediction (NWP) models, which are prone to systematic and random errors. We propose a method that integrates a convolutional neural network (CNN) with an ensemble of low-resolution NWP models (40-km horizontal resolution) to produce high-resolution (5-km) surface temperature forecasts with lead times extending up to 5.5 days (132 h). First, CNN-based post-processing (bias correction and spatial downscaling) is applied to individual ensemble members to reduce systematic errors and perform downscaling, which improves the deterministic forecast accuracy. Second, this member-wise correction is applied to all 51 ensemble members to construct a new high-resolution ensemble forecasting system with an improved probabilistic reliability and spread–skill ratio that differs from the simple error reduction mechanism of ensemble averaging. Whereas averaging reduces forecast errors by smoothing spatial fields, our member-wise CNN correction reduces error from noise while maintaining forecast information at a level comparable to that of other high-resolution forecasts. Experimental results indicate that the proposed method provides a practical and scalable solution for improving medium-range temperature forecasts, which is particularly valuable for use in operational centers with limited computational resources.

SIGNIFICANCE STATEMENT

Reliable temperature forecasts lasting more than five days are vital for planning and public safety; however, current forecasting methods rely on low-resolution numerical weather prediction models that frequently misestimate temperatures. We developed a method that employs artificial intelligence to correct each member of a set of forecasts (an “ensemble”), thereby reducing systematic and random errors. This method provides more accurate, reliable, and spatially detailed ensemble forecasts than the model’s original output while remaining feasible under limited computing resources. This method can help communities better prepare for temperature-related risks by making accurate forecasts accessible to operational centers with limited resources.

## 1. Introduction

Surface temperature is an important meteorological variable that influences sectors such as agriculture, transportation, tourism, and public health (Frisvold and Murugesan 2013; Xia et al. 2013; Wang et al. 2023; Yin et al. 2024). Decision making in these sectors primarily relies on the temperature forecasts of numerical weather prediction (NWP) models. Although improvements in model physics, data assimilation, and computational resources have enhanced the predictive accuracy of NWP models (Kawabata et al. 2007, 2011; Bauer et al. 2015; Ikuta et al. 2021), model outputs are prone to systematic and random errors (Stensrud 2007). Systematic errors arise from factors such as inherent model imperfections and limited horizontal resolution of complex topography, whereas random errors are primarily sourced from uncertainties in initial conditions (Lorenz 1969; Palmer 2001; Skamarock 2004).

In most NWP models, a trade-off exists between horizontal resolution and forecast lead time. High-resolution models often shorten the forecast range to prevent the overload of limited computational resources, whereas low-resolution models allow longer forecast lead times. Both systematic and random errors tend to accumulate with decreasing resolution and increasing forecast lead time. Various post-processing techniques have been employed to address these issues. Traditional methods such as the use of model output statistics (Glahn and Lowry 1972) and Kalman filtering (KF) (Homleid 1995; Anadranistakis et al. 2004) have been widely used because they effectively reduce systematic errors in NWP forecasts.

Deep learning has recently emerged as a promising post-processing tool for deterministic forecasts, particularly for bias correction and downscaling (e.g., Baño-Medina et al. 2020). In particular, convolutional neural networks (CNNs), which extract hierarchical features using combined convolutional and pooling layers, are widely employed in image recognition (Krizhevsky et al. 2012; Simonyan and Zisserman 2015) and NWP fields to capture spatial error structures. Sayeed et al. (2023) employed CNNs for correcting biases in multiple meteorological variables derived from the Weather Research and Forecasting (WRF) model. Kudo (2022) employed an encoder–decoder CNN architecture to improve the surface temperature forecasts of operational NWP models, successfully resolving the errors in frontal position prediction. Inoue et al. (2024) integrated CNN-based post-processing with KF to improve gridded and point-based surface temperature predictions. Advanced CNN architectures such as U-Net with its skip connections have proven particularly effective for spatially complex variables like precipitation (Grönquist et al. 2021; Hu et al. 2023). These studies indicate that CNNs can effectively reduce systematic errors and downscale NWP forecasts. Beyond CNN architectures, transformer-based models have emerged as a promising alternative for weather forecasting and AI weather models owing to their ability to capture long-range spatial dependencies (Bouallègue et al. 2024; Pathak et al. 2022). However, the computational demand of the self-attention mechanism scales quadratically with the input resolution, which can strain limited computational resources (Bonev et al. 2023; Pathak et al. 2022).

Forecast uncertainty can also be quantified using ensemble prediction methods (e.g., Leith 1974; Toth and Kalnay 1997; Wu et al. 2025), which account for uncertainties in initial and boundary conditions as well as model physics. Ensemble averaging reduces random errors, as demonstrated in the global ensemble forecast dataset produced under the international program *TIGGE* (see https://www.ecmwf.int/en/research/projects/tigge). However, ensemble averaging inherently smooths spatial structures (Swinbank et al. 2016), thereby degrading the representation of fine-scale features and extremes in predictions. Moreover, systematic errors remain even after ensemble averaging, which must be addressed by employing bias-correction techniques (Wang et al. 2018). A prominent category of statistical post-processing for ensembles is ensemble model output statistics (EMOS; Gneiting et al. 2005), which corrects the forecast distribution at each grid point independently. While effective for distributional biases, these point-wise methods cannot explicitly model spatial error structures (Rasp and Lerch 2018). Vannitsem et al. (2021) provided a broad overview of statistical post-processing for ensemble forecasts. Schulz and Lerch (2022) offered a more recent review that focused specifically on the application of machine learning methods.

The Japan Meteorological Agency (JMA) operates the Global Ensemble Prediction System (GEPS; Japan Meteorological Agency 2023), a global-scale ensemble forecast system that supports medium-range weather prediction. The GEPS comprises 51 ensemble members generated by perturbing the initial conditions of a single NWP model to replicate forecast uncertainty. However, owing to its low horizontal resolution, the GEPS cannot capture fine-scale temperature variabilities, particularly in topographically complex regions. Therefore, the GEPS accuracy must be enhanced by employing post-processing techniques such as bias correction and statistical downscaling.

To address the limitations of traditional ensemble post-processing, recent studies have begun to integrate CNNs with ensemble forecasting. For instance, Sha et al. (2022) and Hess and Boers (2022) applied CNNs to ensemble NWP outputs for precipitation forecasting. However, these approaches typically operate on ensemble means or selected summary statistics, which do not explicitly exploit the full variability among individual ensemble members. Consequently, the role of CNN correction applied at the ensemble-member level and its interaction with ensemble averaging remain unclear. In particular, the following knowledge gaps are yet to be well addressed. (i) Which ensemble members should be used as training data for CNNs? (ii) How do CNNs improve the deterministic and probabilistic prediction skills? (iii) What is the relation between ensemble averaging and smoothing effects?

To address these gaps, a unified post-processing framework that integrates CNN-based bias correction and downscaling with ensemble forecasting is employed herein for medium-range surface temperature prediction. First, the CNN is applied to each individual ensemble member to improve its deterministic forecast skill. The set of corrected members is then aggregated to obtain a new and high-resolution ensemble forecast with an improved reliability and spread–skill ratio. An encoder–decoder CNN architecture was adopted owing to its proven success for time-series temperature prediction (Kudo 2022; Inoue et al. 2024), which is less spatially complex than precipitation prediction, and because of its computational efficiency compared with the more intricate U-Net or more costly transformer models, which is crucial for operational use. Using this approach, we aim to answer the following scientific questions:

Q1. How effectively can a CNN-based post-processing framework, applied to individual ensemble members, reduce systematic and random errors in NWP forecasts?

Q2. How does the selection of an ensemble member for training influence the performance of CNN-based post-processing?

Q3. How does CNN-based post-processing improve the reliability and spread–skill ratio, and how does its error correction differ from the smoothing effects of ensemble averaging?

The remainder of this paper is organized as follows. Section 2 describes the considered datasets and preprocessing procedures. The proposed CNN architecture and associated verification metrics are introduced in Section 3. Section 4 outlines the experimental setup, and Section 5 presents the results and their comparison with the results of other models and case studies. Section 6 discusses the implications and limitations of the proposed method, and Section 7 concludes the paper with key findings and future directions.

## 2. Data

### *a. Deterministic NWP Models*

Herein, two deterministic NWP models operated by the JMA are employed as reference for comparison: the global spectral model (GSM; Japan Meteorological Agency 2023) and the mesoscale model (MSM; Japan Meteorological Agency 2023). The GSM is a global hydrostatic model with a horizontal resolution of approximately 20 km in this study. GSM generates forecasts at four daily initialization times: 0000, 0600, 1200, and 1800 UTC. The runs initialized at 0000 and 1200 UTC have a lead time of 216 h, with the runs at 0600 and 1800 UTC having a shorter lead time of 132 h. For all forecasts, output data are provided at 6-h intervals. Herein, GSM forecasts with a lead time of 132 h from 0000 and 1200 UTC are used to ensure consistency with the GEPS dataset.

The MSM with a horizontal resolution of 5 km employing the nonhydrostatic regional NWP model provides high-resolution forecasts. This model is operated eight times per day and provides forecasts with lead times of up to 51 h. Forecasts initialized at 0000 and 1200 UTC are used to maintain temporal consistency with the GEPS and GSM datasets. Owing to its high spatial resolution, the MSM can capture small-scale features such as local temperature variations and precipitation associated with complex terrain. Therefore, the performance of the proposed method is evaluated against MSM output.

Herein, medium-range forecasts are referred to as predictions extending from beyond 3 days up to 10 days (72–240 h) ahead, following standard meteorological convention (World Meteorological Organization 2023). Accordingly, the GSM is classified as a medium-range NWP model, whereas the MSM is primarily used as a short-range reference for validation.

*b. Ensemble NWP Model*

The GEPS provides global-scale forecasts based on a low-resolution ensemble system; it was introduced into the operational system of the JMA on 17 January 2017. Initially, the GEPS comprised 27 ensemble members with a horizontal resolution of 40 km; it was subsequently expanded to 51 members on 30 March 2021. GEPS forecasts are initialized four times per day (0000, 0600, 1200, and 1800 UTC). The maximum lead time depends on the initialization time and, in some cases, the day of the week. For example, the runs initialized at 0600 and 1800 UTC have a maximum lead time of 132 h, whereas the 1200 UTC runs on Tuesdays and Wednesdays have a maximum lead time of 34 days. To ensure consistency with the deterministic models described previously, GEPS forecasts initialized at 0000 and 1200 UTC with lead times up to 132 h are used. The ensemble forecasts involve one control forecast (CF), which generally employs the same physical parameterization schemes as the GSM, but at a lower resolution, and 50 perturbed forecasts (PFs). The perturbations are generated by combining singular vectors (Buizza and Palmer 1995) and the local ensemble transform KF (LETKF; Hunt et al. 2007). Although the operational resolution of the GEPS was upgraded to 27 km on 15 March 2022, 40-km resolution data are employed herein to ensure consistent data usage between training and inference across models.

*c. Post-Processed Baseline*

The JMA applies a KF-based post-processing system for correcting systematic errors in NWP temperature forecasts (Sannohe 2018). The KF system first predicts the 1.5-m air temperature at surface observation stations in Japan via the statistical regression of data derived from nearby NWP grid points. The World Meteorological Organization (2024) recommends placing air-temperature sensors at a screen height between 1.25 and 2 m above ground level. In line with this guideline, JMA adopts 1.5 m as its operational standard. The KF sequentially updates the regression coefficients based on real-time observations. The advantage of this sequential updating is that it allows correction to follow time-varying NWP biases (including seasonal variations and model upgrades), adapt to changes in station environments (e.g., relocations), and accommodate newly established stations without requiring a long-term observational dataset. The resulting point-based forecasts are interpolated onto regular spatial grids via weighted averaging that accounts for topography and distance (Kuroki 2017), yielding bias-corrected temperature forecasts at a horizontal resolution of 5 km.

To establish additional baseline methods for comparison, the KF post-processing framework is applied to both GSM and high-resolution MSM forecasts. Hereafter, forecasts corrected using this framework are denoted with a "–KF" suffix (e.g., GSM–KF and MSM–KF). Although the GSM–KF and MSM–KF systems have limited forecast ranges (84 and 51 h, respectively), they provide effective bias correction and are strong baseline methods for evaluating post-processing methods.

*d. Ground-truth Surface Temperature*

The 1.5-m air temperatures derived from the operational weather distribution products of the JMA are used as ground-truth data (Wakayama et al. 2020). These products provide hourly gridded data at a resolution of 1 km for surface temperature, weather category, and sunshine duration across Japan. Surface temperatures are estimated from the combined observations at more than 900 stations and the gridded climatological normal. This normal is derived using a multiple linear regression model that relates 30-year average temperatures at each observatory, topographic information, and various urban factors. This model enables reliable temperature estimations even in locations lacking direct observations (Wakayama et al. 2020). The accuracy of this dataset is evaluated via cross-validation against actual observations from January 2013 to May 2015. For each grid cell containing an observation station, the estimated temperature is compared with the observed value at that station using a leave-one-out approach, in which the target station is excluded from estimations. The bias and root mean square error (RMSE) are approximated as 0.01 and 1.19 K, respectively, indicating the high accuracy of the dataset (Japan Meteorological Agency 2016). These values represent the nationwide average across Japan, including mountainous regions, where both interpolation errors and extreme temperature variability are relatively larger compared with those over plains. Within the common temperature range (−5°C–25°C), the RMSE is approximately 1 K and is below 1 K over the plains considered in case studies.

Herein, the estimated temperatures are gridded onto a 5-km mesh by simple averaging to ensure consistency with the 5-km resolution of the GSM–KF system. Given that the temperature field is derived from a process of climatological regression and spatial interpolation, any error induced by this conservative aggregation method should not be large enough to alter our conclusions. These 5-km-averaged estimated surface temperatures (ESTs) are treated as the target (ground truth) for evaluating gridded temperature predictions.

*e. Summary of Datasets*

Several datasets provided by the JMA are used as inputs and references for evaluating the proposed method. The GEPS is the main input dataset for CNN-based bias correction and downscaling. The GSM provides high-resolution deterministic forecasts based on generally the same physical parameterization schemes as those of the GEPS. For comparative evaluations, the MSM is used as a high-resolution reference dataset and the GSM–KF and MSM–KF systems are used as high-resolution post-processed reference datasets. The 5-km gridded EST dataset is used as the ground truth for both training and evaluation. This dataset extends from 0000 UTC 17 January 2017 to 1200 UTC 31 December 2022 at 12-h intervals and is divided into training, validation, and test periods (Table 1).

| Dataset category | Time period | Data |
| --- | --- | --- |

| Training | 17 January 2017–31 December 2020 | GEPS (40 km), GSM (20 km), and EST (5 km) |
|---|---|---|
| Validation | 1 January 2021–31 December 2021 | GEPS, GSM, and EST |
| Test | 1 January 2022–31 December 2022 | GEPS, GSM, EST, MSM (5 km), GSM–KF (5 km), and MSM–KF (5 km) |

Table 1. Categories and time periods of datasets used for training, validating, and testing the CNN model.

## 3. Methods

### *a. CNN Model*

The CNN-based post-processing framework mitigates the errors introduced by low grid resolutions, terrain-induced biases, and systematic deviations in NWP forecasts.

Figure 1 and Table A1 summarize the CNN architecture and hyperparameters, respectively. The proposed model is based on the encoder–decoder CNN architectures developed by Kudo (2022) and Inoue et al. (2024), which are designed for matched input and output grids to capture spatial patterns in meteorological data derived from NWP models and accurately reconstruct temperature fields. The encoder–decoder structure effectively learns the spatial correlations among meteorological variables, thereby facilitating the refinement of local temperature fields. The CNN architecture comprises convolutional, pooling, and fully connected layers. The rectified linear unit (ReLU) activation functions (Nair and Hinton 2010) are added in the intermediate layers and a sigmoid activation function is added in the output layer to constrain the predictions within a realistic range. Batch normalization is applied to stabilize training and increase convergence.

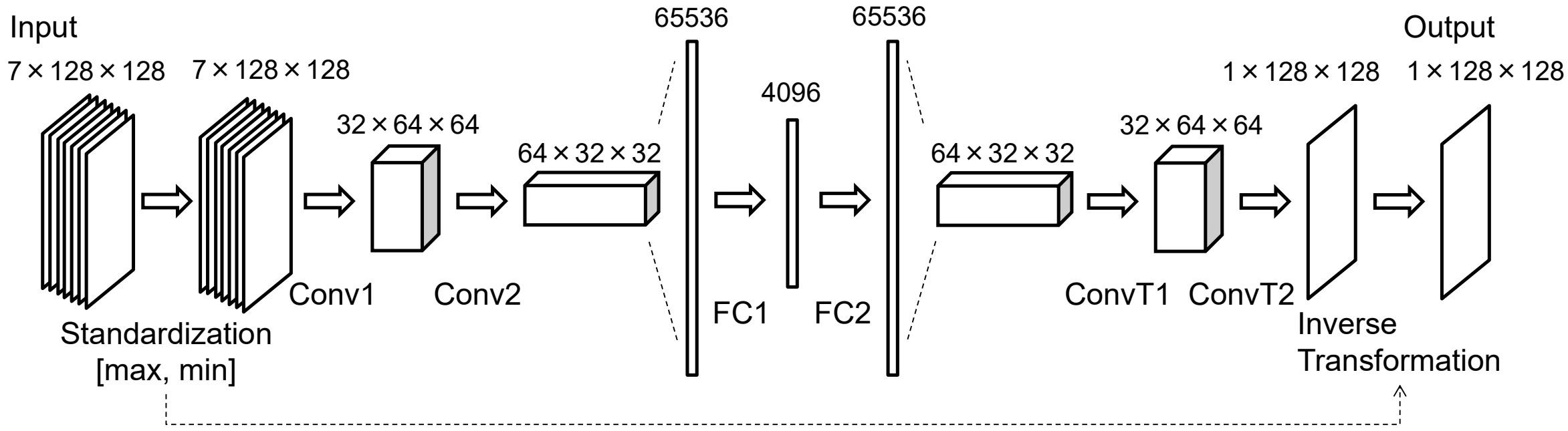

Fig. 1. Schematic of the CNN (reprinted from our previous work: Fig. 1. in Inoue et al. 2024). Conv1, Conv2, and other operational units are detailed in Table A1.

The CNN model is trained using seven meteorological variables obtained from NWP outputs: surface temperature, temperatures at 975, 925, and 850 hPa, mean-sea-level pressure (MSLP), and the zonal (U) and meridional (V) components of the surface wind. These variables are selected because they influence the surface temperature via thermodynamic and dynamical processes. Additional inputs such as time-of-day information and static predictors such as topographic information are not explicitly included as CNN inputs because surface temperatures exhibit strong diurnal variability and elevation-related effects manifest as systematic errors. Therefore, these factors are assumed to be implicitly learned from temperature fields, similar to those in previous studies (Kudo 2022; Inoue et al. 2024). All input variables are normalized to the [0, 1] range to improve training stability and ensure consistent scaling across different meteorological parameters. For this standardization, the spatial maximum ($\phi_{\max}$) and minimum ($\phi_{\min}$) values are derived from the input NWP data at each target time following the approach of Kudo (2022). The specific min–max normalization is as follows:

$$\phi' = \frac{\phi - \phi_{\min}}{\phi_{\max} - \phi_{\min}},$$

where $\phi$ and $\phi'$ denote the original and normalized variables, respectively. This instance-specific normalization method yields a normalized variable distribution that is more balanced and adaptive to the daily dynamic range of meteorological variables within the [0, 1] range compared with that based on a single global maximum and minimum derived from the entire training dataset. To account for extreme temperatures that fall outside the range predicted by NWP models, the normalization range for temperatures at the surface and 975, 925, and 850 hPa is extended by adding 3 K to the spatial maximum ($\phi_{\max}$) and subtracting 3 K from the spatial minimum ($\phi_{\min}$). For other variables, such as MSLP and wind components, the spatial minimum and maximum at each target time are used without extension. The same range is applied for the inverse transformation of the CNN outputs and for normalizing the EST dataset used as the ground truth. However, this approach may still be unable to predict unprecedented and record-breaking events that fall considerably outside this extended range. Although topographic information is not explicitly used as the CNN input, a land–sea mask is applied so that only land grid points are used to compute the loss function. Thus, the CNN outputs are exclusively evaluated over land areas.

Experiments are conducted over central Japan (Fig. 2), which is chosen for its high population density, diverse topography, and relatively large seasonal variability in surface temperature. This region encompasses both plains and mountainous areas, which makes it an ideal testbed for evaluating the effectiveness of CNN-based post-processing at mitigating terrain-induced temperature biases and improving the spatial detail of forecasts.

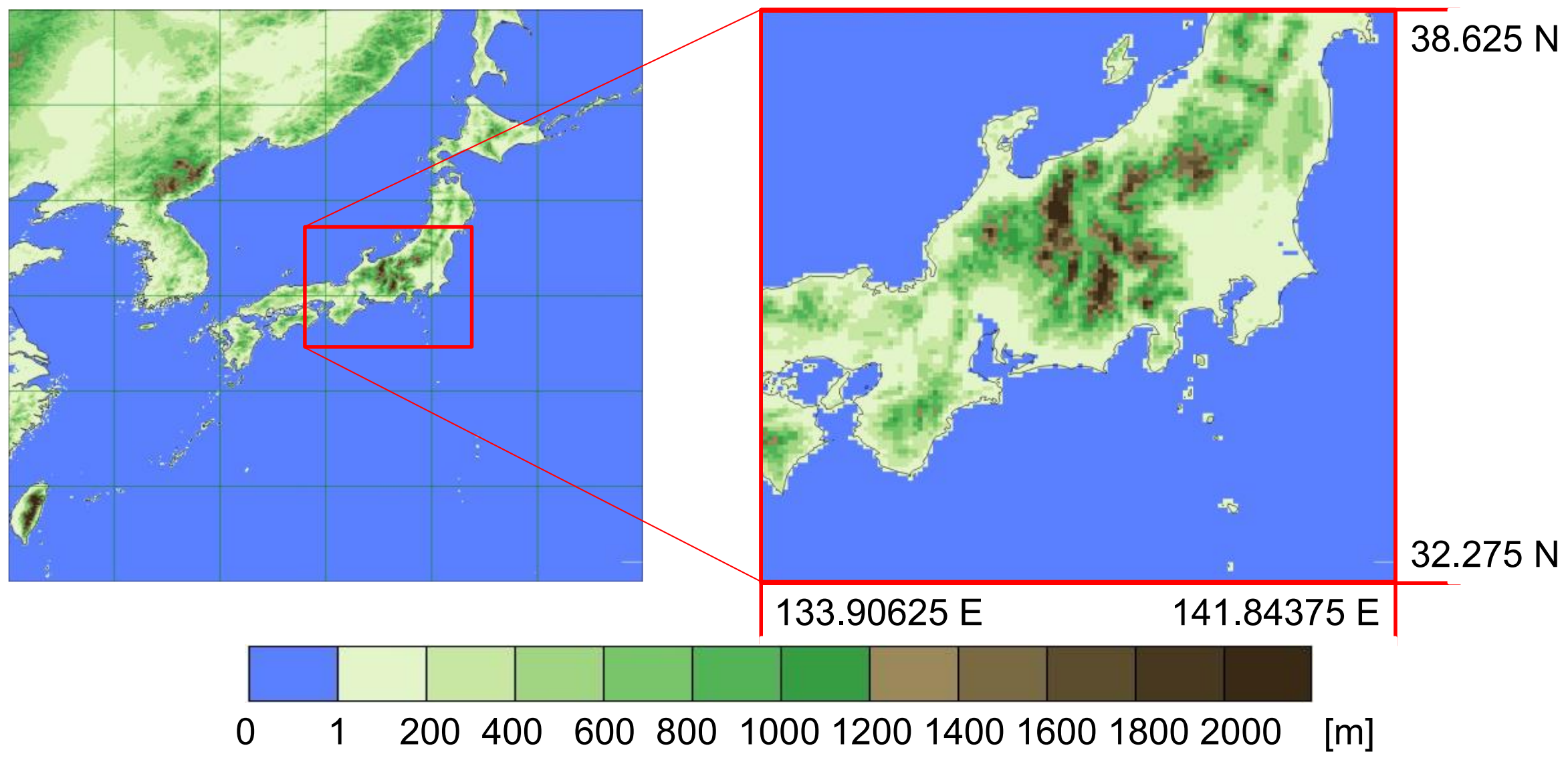

Fig. 2. Islands of Japan located in the Far East (left) and the target domain (right). The color scale indicates elevation. Only the land area is targeted herein.

*b. CNN-based Error Correction and Downscaling for Ensemble Forecasting*

Figure 3 illustrates our post-processing framework, which integrates a CNN into an ensemble forecasting system and comprises distinct training and inference phases. During the training phase, the CNN learns to correct systematic errors and enhance the spatial resolution by mapping the low-resolution NWP outputs to high-resolution gridded observational data. The explanatory variables are the NWP data at horizontal resolutions of 40 km (GEPS) or 20 km (GSM), and the target variable is the 5-km resolution EST. After learning this mapping, the CNN performs bias correction and statistical downscaling by reconstructing high-resolution spatial structures from low-resolution forecasts. In this context, the CNN functions as a downscaling model by capturing various sources of systematic errors introduced by the low resolution of the terrain. It minimizes these errors via training and establishing a data-driven relation, enabling the correction of these systematic errors to enhance the horizontal resolution.

During the inference phase, the trained CNN is applied to the CF and all 50 PFs. Because the CF and PFs share identical model configurations excluding their initial conditions, the error characteristics learned from the CF are assumed to be statistically transferable to the PFs. This assumption is tested in the experiments. This member-wise strategy is adopted because the objective of the present framework is to construct a corrected ensemble forecast. Although some previous studies use the ensemble mean as input, correcting only the ensemble mean in the present framework would yield only a corrected deterministic mean field. This process yields a 51-member CNN-corrected ensemble (GEPS–CNN). Hereafter, CNN-corrected forecasts are appended with the "–CNN" suffix: CF–CNN for the CNN-corrected CF, GSM–CNN for the CNN-corrected GSM, and GEPS–CNN for the ensemble forecast of each individually CNN-corrected GEPS member. Standard ensemble statistics such as the ensemble mean and probabilistic verification scores are then computed from the original GEPS and GEPS–CNN, which allows us to assess how member-wise CNN correction affects the deterministic skill, ensemble spread, and forecast reliability.

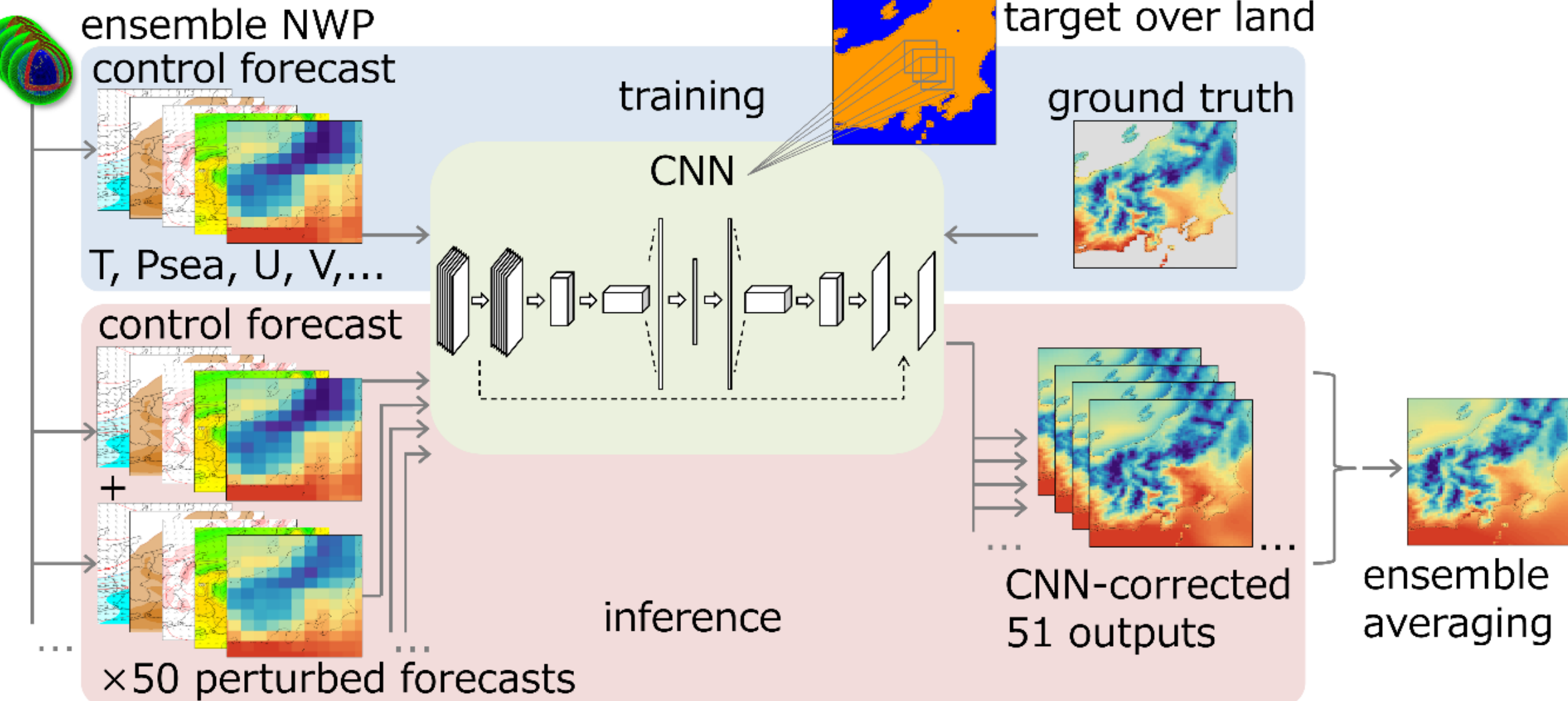

Fig. 3. Schematic of CNN-based post-processing applied to the GEPS. During training (upper panel), the CNN learns mapping parameters from the low-resolution fields of the CF (e.g., temperature, mean-sea-level pressure, and winds) to high-resolution gridded analyses over land. During inference (lower panel), the trained CNN is applied to the CF and all 50 PFs to produce a 51-member CNN-corrected ensemble, from which ensemble statistics such as the ensemble mean are computed.

*c. Ensemble Averaging*

Each ensemble member represents uncertainties arising from initial conditions, boundary conditions, and physical parameterization schemes. Ensemble averaging, which reduces random errors in individual forecasts (Leith 1974), is defined as follows:

$$\bar{f}(t) = \frac{1}{M}\sum_{i=1}^{M} f_i\,(t), \tag{1}$$

where $\bar{f}(t)$ represents the ensemble mean at time $t$, $M$ is the total number of ensemble members, and $f_i(t)$ denotes the forecast of the $i$-th ensemble member. However, the averaging process inherently smooths the spatial fields and loses information on the forecast uncertainty represented by the ensemble.

The ensemble means are appended with "–EM." In particular, GEPS–EM represents the ensemble mean of the GEPS members and CNN–EM denotes the mean of a CNN-corrected GEPS member (GEPS–CNN).

*d. Verification Methods*

The prediction accuracy of the surface temperature is assessed at each grid point within the target area (Fig. 2) from 1 January to 31 December 2022. GEPS forecasts and EST data are used as the input and ground-truth data, respectively. The GEPS and GSM forecasts with horizontal resolutions of 40 and 20 km, respectively, are interpolated to a common 5-km grid using the same bicubic interpolation procedure before subsequent comparisons with the EST data. Forecasts are initialized at 0000 and 1200 UTC and validated at 12-h intervals. For comprehensive evaluation, the verification metrics are organized into three categories: (1) grid-point-based verification, (2) spatial verification, and (3) information content verification.

1) GRID-POINT-BASED VERIFICATION

Herein, forecast performance is evaluated at individual grid points independent of surrounding locations.

*(i) For Deterministic Forecasts*

We employ two primary metrics. The RMSE quantifies the overall magnitude of forecast errors (Wilks 2011):

$$\text{RMSE} = \sqrt{\frac{1}{N}\sum_{n=1}^{N}(F_n(\tau) - O_n(\tau))^2}\,, \tag{2}$$

where $N$ represents the number of grid points and $F_n(\tau)$ and $O_n(\tau)$ denote the forecast and observed temperatures, respectively, at location $n$ and forecast lead time $\tau$. A lower RMSE indicates a higher forecast accuracy.

The mean error (ME) quantifies the average bias between the predicted and observed values:

$$\text{ME} = \frac{1}{N}\sum_{n=1}^{N}(F_n(\tau) - O_n(\tau))\,,$$

where all terms are defined in Eq. (2). An ME close to zero indicates minimal bias, whereas a positive or negative ME indicates systematic overestimation or underestimation, respectively.

*(ii) For Ensemble Forecasts*

To evaluate the ensemble forecast, we use the continuous ranked probability score (CRPS; Hersbach 2000). CRPS is a statistical metric widely used for probabilistic forecast evaluations and is calculated as follows:

$$\text{CRPS}(F(\tau), O(\tau)) = \frac{1}{M}\sum_{i=1}^{M}|f_i(\tau) - O(\tau)| - \frac{1}{2M(M-1)}\sum_{i=1}^{M}\sum_{\substack{j=1\\ j\neq i}}^{M}|f_i(\tau) - f_j(\tau)|\,,$$

where $F(\tau)$ is the ensemble forecast and $f_i(\tau)$ and $f_j(\tau)$ are the forecasts of the $i$ and $j$-th ensemble members, respectively, at forecast lead time $\tau$, $M$ is the total number of ensemble members, and $O(\tau)$ is the observed value. Low CRPSs indicate high probabilistic forecast skill. CRPS is computed at each grid point and then averaged over all grid points to obtain a single score at each forecast lead time $t$. Although the standard unweighted CRPS summarizes overall performance, aggregating CRPS across locations with differing error characteristics can mask conditional performance issues, as discussed by Bolin and Wallin (2023).

Next, we assess the ensemble's internal characteristics. The ensemble spread represents the forecast uncertainty and is computed as the standard deviation across ensemble members:

$$S(t) = \sqrt{\frac{1}{M-1}\sum_{i=1}^{M}(f_i(t) - \bar{f}(t))^2}\,, \tag{3}$$

where $S(t)$ represents the ensemble spread at time $t$ and $\bar{f}(t)$ is the ensemble mean defined by Eq. (1). The spread–skill ratio (SSR) is the ratio of the ensemble spread (Eq. 3) to the RMSE (Eq. 2) of the ensemble-mean forecast, and an SSR close to 1 indicates a well-calibrated ensemble. The SSR does not explicitly account for observational uncertainty in the EST. However, because both the original and post-processed ensembles are verified against the same EST, this observational error affects both systems equally and does not impact any comparisons between them.

2) SPATIAL VERIFICATION

While point-based metrics like RMSE are valuable, they cannot evaluate the spatial attributes of a forecast, such as the location, shape, and texture of weather features. To address this limitation and to validate our claims that the CNN enhances spatial detail, we utilize the following metrics for spatial verification.

*(i) For Deterministic Forecasts*

The use of the fractions skill score (FSS; Roberts and Lean 2008) is a prominent example of neighborhood-based methods for spatial verification (e.g., Gilleland et al. 2009; Dorninger et al. 2018). FSS is defined as

$$\mathrm{FSS}(\tau) = 1 - \frac{\sum_{n=1}^{N_x}\sum_{l=1}^{N_y}\left(F(\tau)_{(s)n,l} - O(\tau)_{(s)n,l}\right)^2}{\sum_{n=1}^{N_x}\sum_{l=1}^{N_y}\left(F(\tau)^2{}_{(s)n,l} - O(\tau)^2{}_{(s)n,l}\right)},$$

where $F(\tau)_{(s)n,l}$ and $O(\tau)_{(s)n,l}$ represent the fractional coverage of grid points exceeding a specific threshold within a neighborhood of size $s \times s$ centered at grid point $(n, l)$ for the forecast and observation, respectively, and the summation is over all $N_x \times N_y$ grid points in the domain. FSS compares the spatial fraction of grid points exceeding a specific threshold between the forecast and observation. This approach mitigates the “double-penalty” problem of grid-point-based metrics. A score of 1 indicates a perfect forecast, and scores greater than 0.5 are generally considered useful.

*(ii) For Ensemble Forecasts*

Graphical tools are essential for diagnosing ensemble reliability. Our primary diagnostic tool is the rank histogram (Hamill 2001), which assesses statistical consistency (i.e., calibration). A flat rank histogram indicates a statistically reliable ensemble, whereas a U-shaped (under-dispersive) or dome-shaped (over-dispersive) distribution reveals under- or over-dispersive tendencies, respectively. However, Hamill (2001) warned that a single domain-aggregated rank histogram can mask conditional biases. Therefore, following the standard diagnostic approach of stratification (e.g., Wilks 2011), we compute rank histograms stratified by physically meaningful conditions. To diagnose potential topography-dependent biases, we compute rank histograms stratified by mountainous areas (elevation ≥ 200 m) and plain areas (elevation < 200 m) (Fig. 2). This approach is consistent with the concept of conditional rank histograms (Bröcker 2008; Siegert et al. 2012) and with the broader philosophy of examining specific structured characteristics when assessing multivariate forecast behavior (Allen et al. 2024).

For a comprehensive evaluation of the ensemble's spatial characteristics, we employ two proper scoring rules that assess multivariate performance and structural fidelity. The energy score (ES; Gneiting and Raftery 2007) is a multivariate generalization of the CRPS that can be used to measure the statistical distance between the forecast ensemble distribution and observation field:

$$\mathrm{ES}\big(F(\tau), O(\tau)\big) = \frac{1}{M}\sum_{i=1}^{M}\|f_i(\tau) - O(\tau)\|^{\beta} - \frac{1}{2M(M-1)}\sum_{i=1}^{M}\sum_{\substack{j=1\\ j\neq i}}^{M}\big\|f_i(\tau) - f_j(\tau)\big\|^{\beta},$$

where $O(\tau) \in \mathbb{R}^m$ is the $m$-dimensional observation vector and $f_i(\tau)$ is the corresponding forecast vector for the $i$-th ensemble member. The term $\|\quad\|$ is the Euclidean norm. The exponent $\beta \in (0, 2)$ controls the weight given to larger errors. Here, we use the standard choice $\beta = 1$, which preserves the physical units of the predictand (K) and avoids undue emphasis on large errors.

While the ES assesses the overall multivariate performance, the variogram score (VS; Scheuerer and Hamill 2015) assesses the realism of the spatial structure by comparing observed pairwise differences between grid points with the corresponding ensemble-mean pairwise differences:

$$\mathrm{VS}\big(F(\tau), O(\tau)\big) = \sum_{n=1}^{m}\sum_{l=1}^{m} w_{nl}\left(|O_n(\tau) - O_l(\tau)|^p - \frac{1}{M}\sum_{i=1}^{M}\big|f_{i,n}(\tau) - f_{i,l}(\tau)\big|^p\right)^2,$$

where $O_n(\tau)$ and $f_{i,n}(\tau)$ are the observed value and forecast value of the $i$-th member at grid point $n$, respectively. The indices $n$ and $l$ run over the $m$ grid points of the spatial field. We set $w_{nl} = 1$ for all grid-point pairs. For the VS, we use $p = 1$, which yields a moderate sensitivity to the magnitude of pairwise differences. Lower values indicate better agreement in spatial structure.

### 3) Information Content Verification

Forecast and observation anomalies are calculated with respect to a daily-hourly climatology derived from the ground-truth data (EST) over the training period (2017–2020) and smoothed using a 15-day centered moving average. These anomalies are then used to investigate the relation among forecast skill, variance, and correlation by computing the forecast information (FI), noise error (NE), and anomaly correlation coefficient (ACC) following the work of Bonavita and Geer (2025; hereafter BG25). The relations among these quantities are represented geometrically using the triangle framework proposed by Feng et al. (2024; hereafter FTZP24) and BG25 (Fig. 4). In this framework, the sides correspond to the standard deviations of analysis anomaly (SDAV), the standard deviation of forecast anomaly (SDAF), and the standard deviation of error. The SDAF, which is the length of the forecast-anomaly vector, represents the forecast activity. The covariance between the forecast and observation anomalies is denoted by $p$, and the angle $\theta$ between the two anomaly vectors defines ACC as follows.

$$\mathrm{ACC} = \cos\theta = \frac{p}{\mathrm{SDAF} \times \mathrm{SDAV}}.$$

FI and NE are derived from the projection of the forecast-anomaly vector onto the observed anomaly vector. FI quantifies the useful (correlated) component of forecast variance, whereas NE represents the uncorrelated (noise) component:

$$\mathrm{FI} = \frac{p}{\mathrm{SDAF}^2}, \qquad \mathrm{NE} = \mathrm{SDAF}\sqrt{1 - \mathrm{ACC}^2}.$$

The information error (IE), corresponding to the unexplained component of the observed anomaly variance, is given by

$$\mathrm{IE} = |1 - \mathrm{FI}| \times \mathrm{SDAV}.$$

FI, representing the fraction of the covariance $p$ explained by the variance of observed anomalies, and NE together correspond to an orthogonal decomposition of the forecast variance into information and noise components (Fig. 4). A forecast system with high FI and low NE retains a large proportion of the observed variability and thus exhibits high predictive skills.

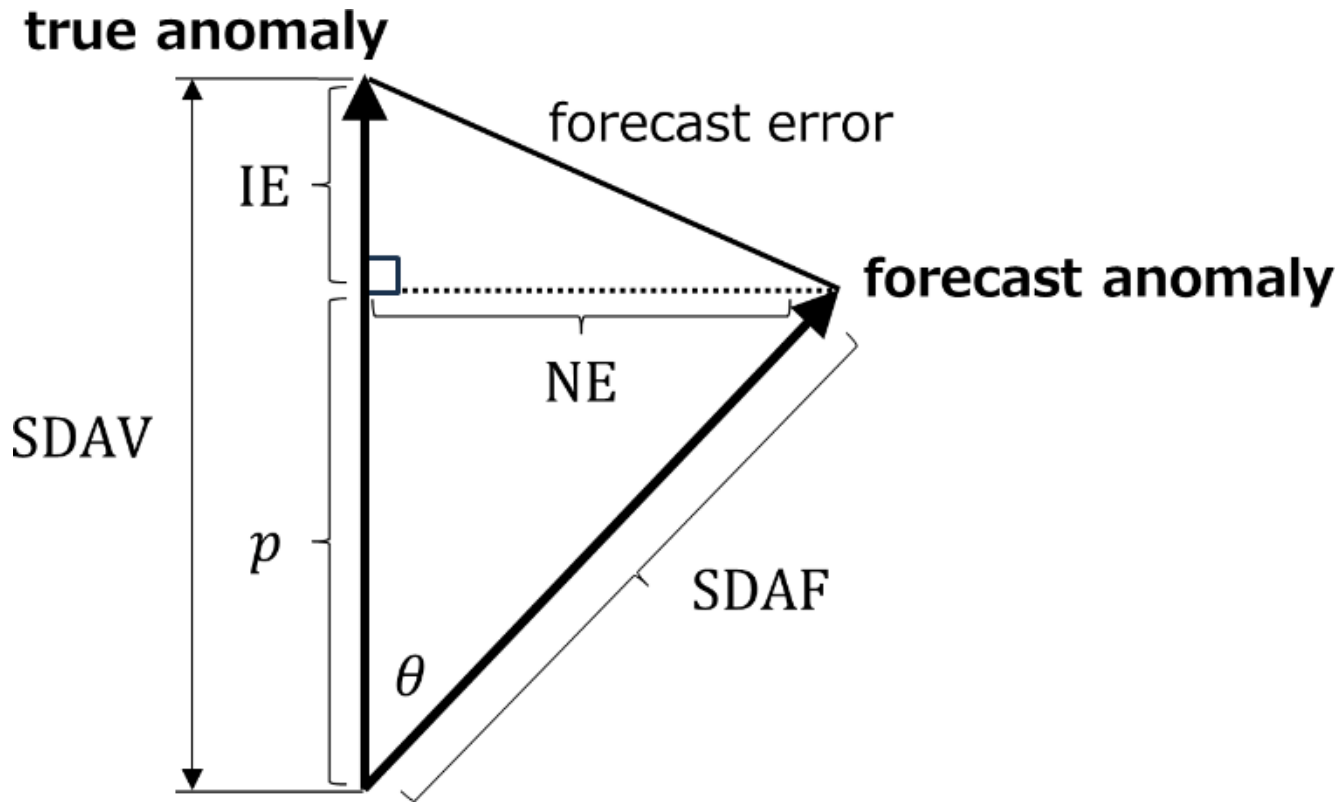

Fig. 4. Geometric representation of the relations between true and forecast anomalies used in the definitions of FI, NE, and ACC (Feng et al. 2024; Bonavita and Geer 2025).

## 4. Experiments

### *a. Deterministic Forecast Correction*

To evaluate the performance of CNN-based post-processing, a CNN is trained on a CF from the training period. The CNN is then applied to a CF from the test period during the inference phase, which yields CF–CNN. The corrected forecasts in CF–CNN are compared with those in the baseline GSM–KF and MSM–KF systems at the same 5-km horizontal resolution. CF–CNN is also compared with the forecasts from the operational NWP models (GSM and MSM) used as inputs for the GSM–KF and MSM–KF systems. Notably, the MSM has a native horizontal resolution of 5 km even before KF post-processing.

An additional CNN is trained using the GSM outputs and applied to the GSM dataset during inference, yielding GSM–CNN. By comparing GSM–CNN with CF–CNN, which employ the same physical parameterization schemes as the GSM but have different horizontal resolutions (40 km for CF vs. 20 km for GSM), the performance of CNN-based post-processing at different input resolutions can be evaluated. The abbreviations used in Sections 4a are summarized in Table 2.

| Abbreviation | Description | Horizontal resolution |
|---|---|---|
| CF | Control forecast of the GEPS | 40 km |
| GSM | Global spectral model, which shares the same physical parameterization schemes as the GEPS | 20 km |
| MSM | Mesoscale model (used as short-range reference) | 5 km |
| GSM–KF | GSM corrected with a Kalman filter (used as post-process reference) | 5 km |
| MSM–KF | MSM corrected with a Kalman filter (used as post-process reference) | 5 km |
| CF–CNN | CF corrected with a CNN | 5 km |
| GSM–CNN | GSM (different horizontal resolution from CF) corrected with a CNN | 5 km |

Table 2. Descriptions of the abbreviations used in Section 4a.

*b. Member-based Training Comparison*

To determine whether training the CNN on the CF is optimal, we compare CNNs trained on three data sources: the CF, one PF (PF10; the 10th member is selected as a representative example), and the GEPS–EM. The rationale for this experiment is our hypothesis that the CNN primarily learns to correct common error components (e.g., biases from model resolution and topography) shared across all members rather than the unique member-specific errors arising from perturbations in the initial conditions. If this hypothesis holds true, then the final forecast skill should be largely insensitive to which member is used for training.

Let $M_{train}$ denote a training data source from the set $\{$CF, PF10, GEPS–EM$\}$. The abbreviations for the resulting models and forecasts are defined in Table 3. Each trained model (CNN ($M_{train}$)) is applied to both CF and PF (CF–CNN ($M_{train}$) and PF–CNN ($M_{train}$)), and their ensemble means are computed (CNN ($M_{train}$)–EM). To ensure a consistent comparison, the same CNN architecture and training settings are used for all configurations.

| Abbreviation | Description | Horizontal resolution |
|---|---|---|

| CF, PF, GEPS–EM | Respective data sources ($M_{train}$) used for training a CNN | 40 km |
|---|---|---|
| CF–CNN ($M_{train}$) | CF corrected with the CNN trained using $M_{train}$ | 5 km |
| PF–CNN ($M_{train}$) | PF corrected with the CNN trained using $M_{train}$ | 5 km |
| GEPS–CNN ($M_{train}$) | Ensemble forecast of GEPS (CF and 50 PF) corrected with the CNN trained using $M_{train}$ | 5 km |
| CNN ($M_{train}$)–EM | Ensemble mean of GEPS–CNN ($M_{train}$) | 5 km |

Table 3. Descriptions of the abbreviations used in Section 4b.

*c. Ensemble Application*

A CNN trained using the CF is applied to all 51 GEPS members, including one CF and 50 PFs. To compare the overall performance of the two ensemble forecasting systems before and after CNN correction, the ensemble mean of the original GEPS forecasts (GEPS–EM) is compared with that of the CNN-corrected forecasts (CNN–EM). The ensemble-mean and member-wise results are compared with those from the deterministic GSM and MSM models, GSM–KF and MSM–KF baselines, and CNN-corrected GSM forecasts (GSM–CNN).

## 5. Results

*a. Deterministic Forecast Correction*

To address Q1, we first examine how CNN-based post-processing applied to individual ensemble members reduces systematic and random errors in deterministic forecasts. Figure 5 shows the verification results of deterministic forecasts before and after post-processing from both grid-point and spatial perspectives. Regarding the grid-point metrics, before post-processing, higher-resolution models exhibit smaller RMSEs (40 km CF > 20 km GSM > 5 km MSM) (Fig. 5a). After post-processing, all forecasts evaluated at 5-km horizontal resolution show reduced RMSEs compared with their original systems. CNN-based post-processing achieves a substantial average improvement of 1.2 K (46%) for CF–CNN relative to the original CF and 0.99 K (39%) for GSM–CNN relative to the original GSM (Fig. 5a). In contrast, KF-based correction (GSM–KF) provides a smaller yet still notable average relative improvement of 0.76 K (29%) relative to the original GSM. Notably, CNN-based post-processing (CF–CNN, GSM–CNN) consistently outperforms KF-based post-processing in reducing both the RMSE and ME (Figs. 5a and b) throughout the forecast range examined. As shown in Fig. 5b, while the KF-based correction does reduce the biases of the raw GSM and MSM models, their original biases are substantial. As a result, the post-processed GSM–KF and MSM–KF forecasts have a residual positive bias. In contrast, the CNN-based correction proves powerful enough to reduce the ME to within ±0.05 K for both CF–CNN and GSM–CNN, which effectively corrects the initial bias regardless of the input model. Moreover, no statistically significant difference is found between CF–CNN and GSM–CNN in either RMSE or ME despite the different original horizontal resolutions of the input forecasts (40 km for the CF and 20 km for the GSM). We further evaluate the FSS at a 48-h lead time, which corresponds to the maximum forecast range of the operational MSM, by using thresholds defined by the mean ± one standard deviation of the domain-averaged temperature during each verification period. In summer (Fig. 5c), the FSS of the KF-corrected models peaks at small to medium scales before declining. This indicates that at larger scales, spatial errors in the overall area and spatial pattern of the event become more dominant than local position errors. The KF method, which applies corrections at observation sites and interpolates the correction increments onto the model grid, cannot directly learn or correct such spatially coherent error structures. In contrast, the FSS for the CNN-corrected models does not decline, which demonstrates their ability to learn and reproduce spatially coherent and realistic spatial patterns from the training data. In winter (Fig. 5d), the CNN-corrected forecasts also maintain the highest FSS over various window sizes. These results confirm that the superiority of the CNN-corrected models extends beyond a grid-point error reduction to an improvement in spatial forecast skill and that this advantage is robust with respect to both the forecast lead time and spatial scale.

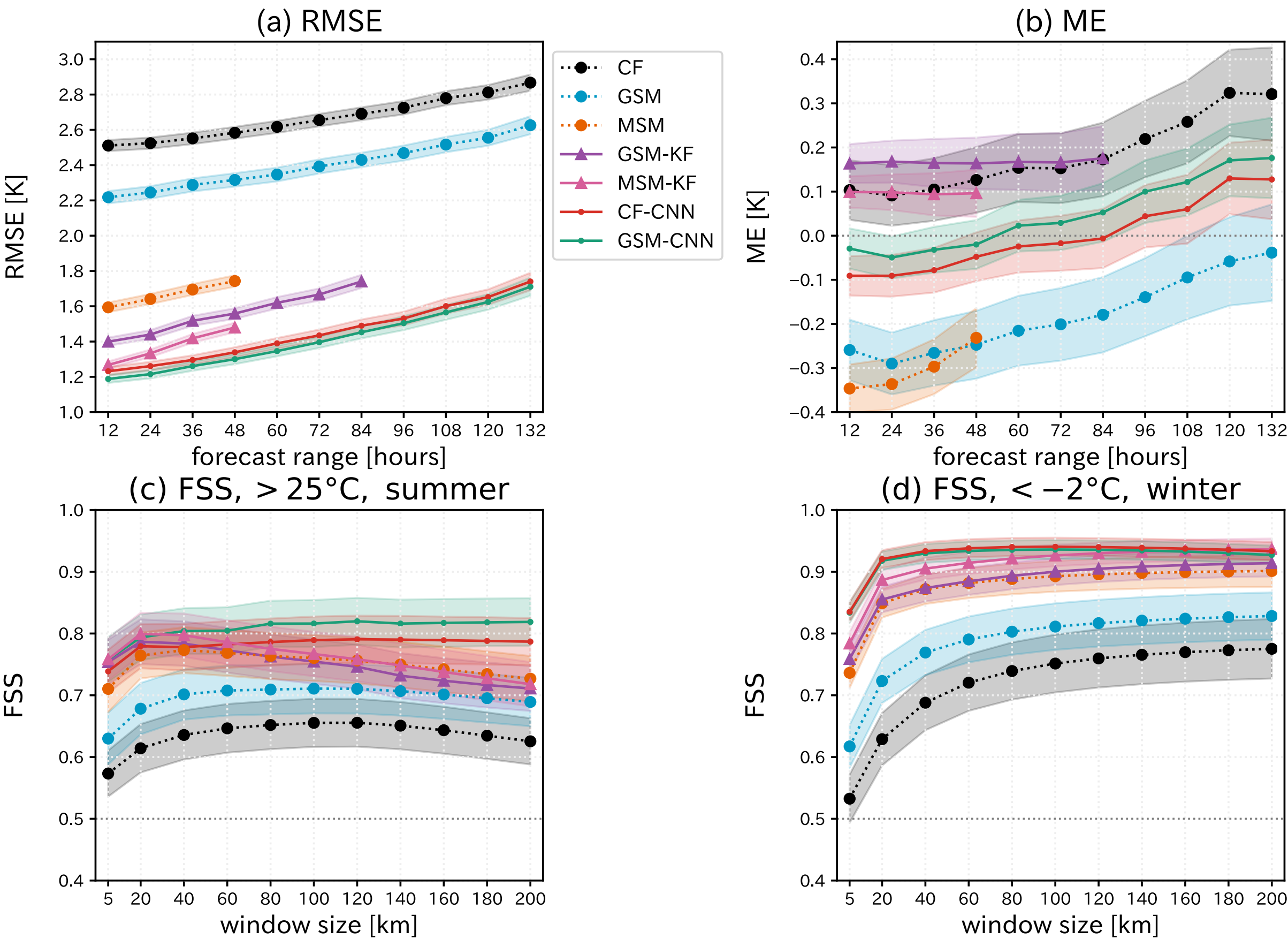

Fig. 5. (a) RMSEs and (b) MEs as functions of the forecast range. (c) FSS for temperatures exceeding 25°C in summer (July–September) and (d) FSS for temperatures falling below −2°C in winter (December–February) as a function of the window size at a lead time of 48 h. All scores are for CF, GSM, MSM, GSM–KF, MSM–KF, CF–CNN, and GSM–CNN, averaged over the 1 January–31 December 2022 period. The shaded areas represent the 95% confidence intervals.

*b. Member-based Training Comparison*

To address Q2, we compare CNNs trained on three data sources: the CF, PF10, and GEPS–EM. Before examining how the choice of training member affects the final ensemble, we first assess how CNN-based post-processing improves the individual deterministic components (i.e., the CF and PFs). As shown in Fig. 6a, CNN-based correction yields substantial improvements over the raw forecasts. For example, the CF–CNN achieves an average RMSE reduction of approximately 1.2 K (46%) relative to the original CF.

After CNN-based post-processing, the CF and PFs show substantial improvement, and their lead time–averaged RMSEs are reduced by approximately 1.2 K. The initial RMSE difference between the raw CF and raw PFs (gray line) is approximately 0.14 K. After correction, this gap between the CF–CNN and PF–CNN increases slightly to about 0.20 K, although this widening is minor relative to the substantial overall error reduction. This is possibly because the CNN primarily removes error components common to all ensemble members such as biases associated with model resolution, topographic information, and physical parameterization schemes, but perturbation-dependent differences among members are not fully eliminated.

Importantly, the final corrected ensemble product is largely insensitive to the choice of training data source. As shown in Fig. 6, both the deterministic skill of the corrected ensemble mean and the probabilistic skill of the full corrected ensemble are very similar regardless of which member is used for training. In particular, the average CRPS is reduced by approximately 0.8 K (a 47% improvement) relative to the original GEPS (Fig. 6b). No statistically significant differences are found among the three CNNs. These results indicate that CNN-based post-processing is robust to the choice of training member; therefore, the CF is used as the training data in all subsequent experiments for simplicity.

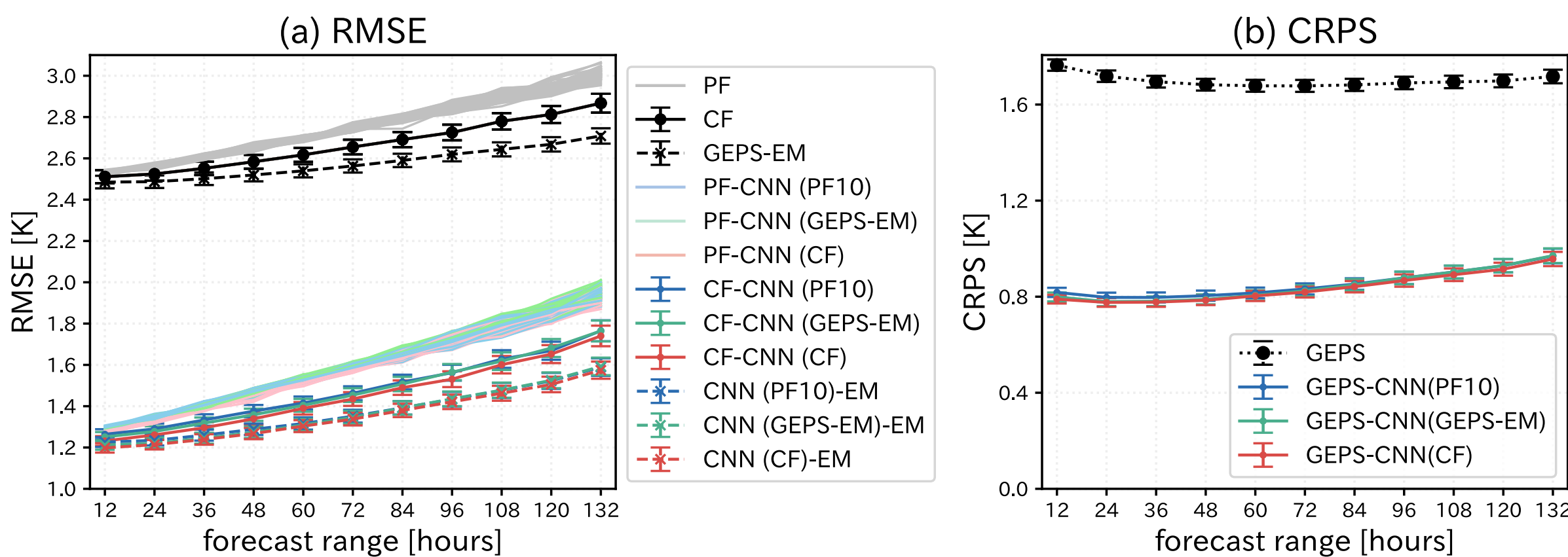

Fig. 6. (a) Average RMSEs and (b) CRPSs of the models at different forecast lead times from 1 January to 31 December 2022. Error bars represent the 95% confidence intervals. Gray-, pink-, light-blue-, and light-green-shaded regions are the visual aggregations (thin lines) of 50 ensemble members (PF and PF–CNN (CF/PF10/GEPS–EM), respectively). The CRPSs are calculated from all GEPS members and the ensembles corrected by each CNN.

*c. Ensemble Application*

To address the first part of Q3, we evaluate how member-wise CNN-based post-processing affects probabilistic skill, calibration diagnostics, and multivariate spatial performance of the final ensemble forecast, GEPS–CNN. The overall probabilistic skill as assessed by the point-wise CRPS is shown in Fig. 6b. The GEPS–CNN consistently outperforms the original GEPS, which indicates an improvement in forecast quality. Figure 7 helps diagnose the source of this improvement. While the CNN slightly reduces the ensemble spread, this reduction is modest at 0.1–0.2 K depending on the lead time. This is far outweighed by the substantial decrease in RMSE of approximately 1.2 K (Fig. 7a). It is this disproportionate improvement in skill relative to the small change in spread that drives the SSR closer to 1.0 (Fig. 7b).

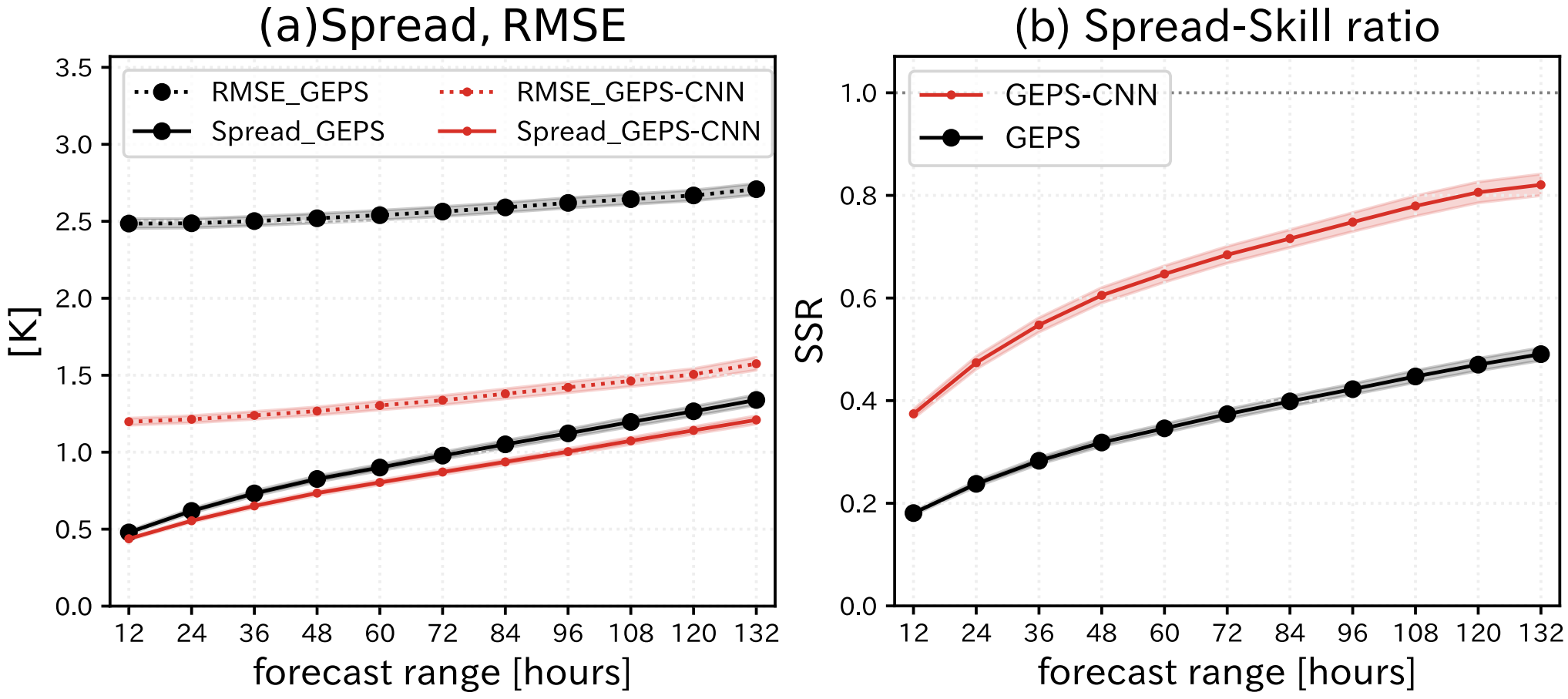

Fig. 7. (a) Ensemble spread and RMSE and (b) SSR based on all 51 ensemble members of the GEPS and GEPS–CNN at each forecast lead time. The experimental period is from 1 January to 31 December 2022. Error bars represent the 95% confidence intervals.

Conditional calibration diagnostics are assessed using rank histograms stratified by elevation (Fig. 8). In the original GEPS, the histograms show strong excess frequency at the extreme ranks, which indicates that observations frequently fall outside the ensemble range. In addition, the asymmetry of the extreme ranks suggests opposing topography-dependent biases with negative bias in mountainous areas and positive bias in plain areas. After CNN-based post-processing, the excess frequency at both extremes is substantially reduced in each stratum and the histograms become flatter. This indicates that the GEPS–CNN better captures the range of observed outcomes while also mitigating the dominant topography-dependent conditional biases.

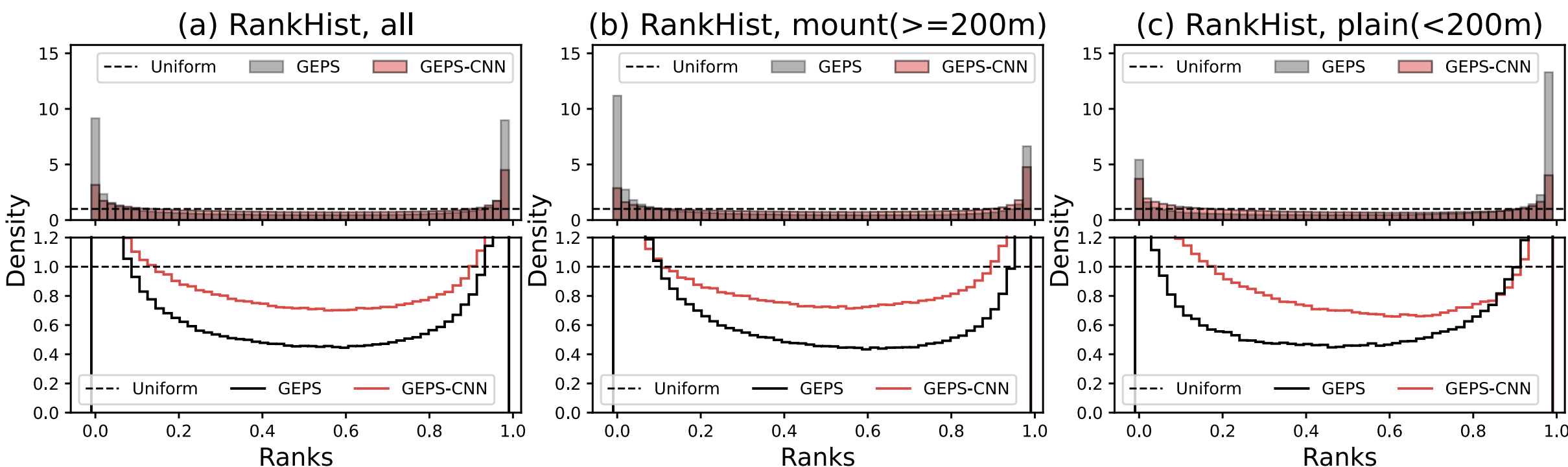

Fig. 8. Rank histograms for the GEPS and GEPS–CNN as calculated for (a) the entire domain, (b) mountainous areas (elevation ≥ 200 m), and (c) plain areas (elevation < 200 m). The dashed line indicates a perfectly calibrated ensemble.

To complement the rank-histogram and SSR diagnostics, Fig. 9 presents the ES and VS for multivariate spatial verification. The ES is a multivariate generalization of the CRPS and is reduced by more than 50%, which indicates an improvement in the overall multivariate performance. The VS, which evaluates the realism of the spatial structure, is reduced by approximately 75% relative to the original GEPS. Together, these results support the conclusion that CNN-based post-processing improves not only pointwise forecast errors but also the realism of the spatial field.

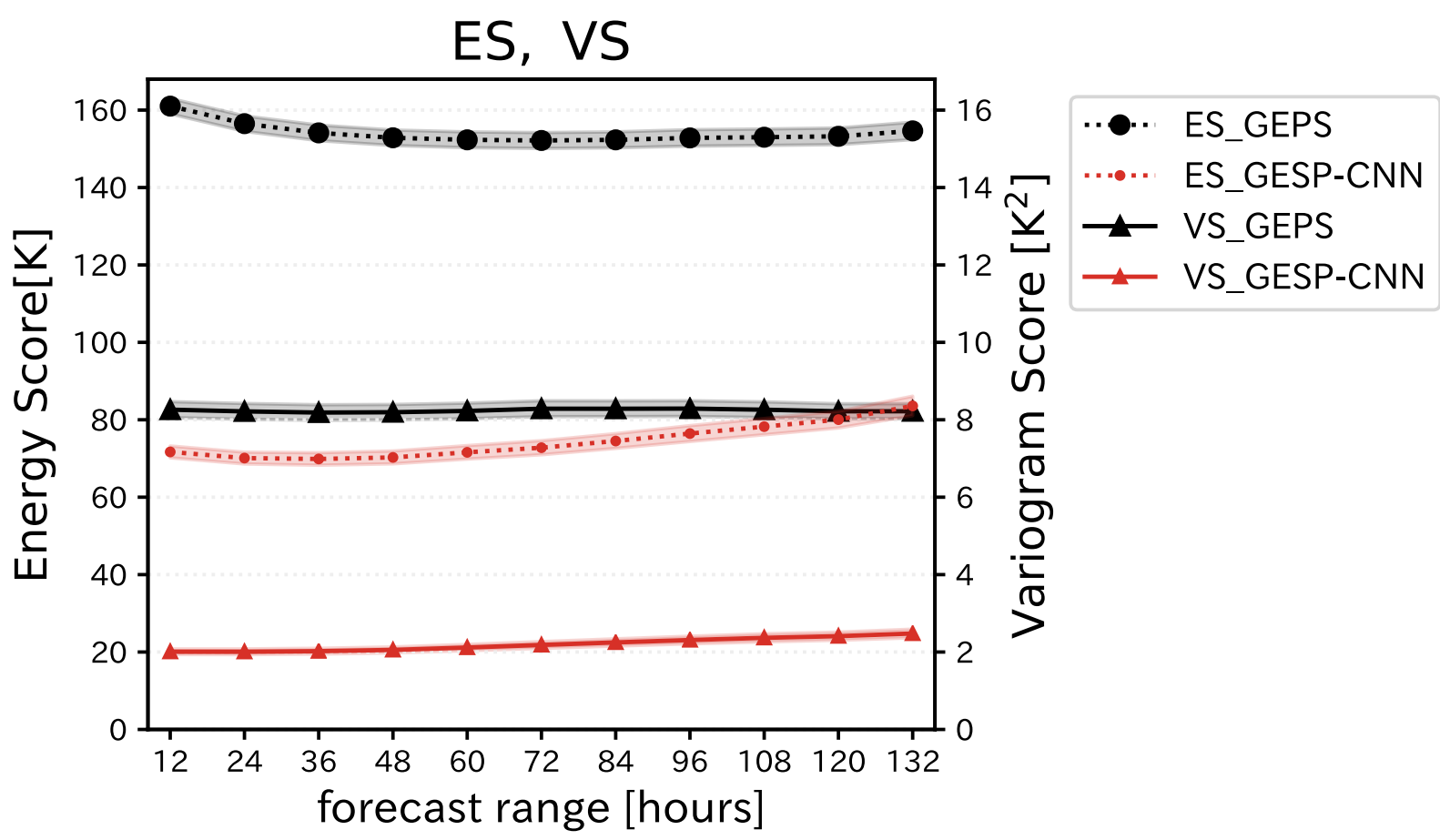

Fig. 9. ES and VS for the GEPS and GEPS–CNN averaged over the experimental period. Lower scores indicate better performance.

*d. Case Studies*

1) Forecast Improvement over Complex Terrain

Here, illustrative examples are presented of forecast behavior over complex terrain for a single lead time of 84 h, which corresponds to the maximum forecast range of the operational GSM–KF. The CF shows a less spatially detailed temperature field because of its 40-km horizontal resolution (Fig. 10a), whereas the CF–CNN shows a more spatially detailed temperature field (Fig. 10b) that corresponds well with the 5-km topography (Fig. 10c), even over mountainous regions with complex terrain.

An elevation-corrected baseline forecast (CF–Hcorr) is constructed by adjusting the CF temperature at each grid point using a typical environmental lapse rate of $-6$ K $\mathrm{km}^{-1}$ (e.g., Dodson and Marks 1997) based on the difference between the model orography and analysis elevation. Compared with the original CF, which has a domain-averaged RMSE of 2.9 K and an ME of –0.84 K, this simple correction reduces the RMSE from 2.9 to 2.3 K but makes the ME more negative from –0.84 to –1.2 K. This indicates persistent negative biases, particularly over the northeastern mountainous region (Fig. 10d). In contrast, the CF–CNN (Fig. 10e) substantially reduces these terrain-dependent biases by reducing the RMSE from 2.9 to 1.4 K and shifting the ME from –0.84 to –0.24 K. Compared with the 20-km GSM forecast (not shown), the GSM–KF forecast shifts the ME from –1.18 to 0.23 K and reduces the RMSE from 2.5 to 1.3 K. However, positive biases associated with elevation remain in the GSM–KF forecast (Fig. 10f). Overall, the CF–CNN produces smaller errors than the CF–Hcorr, which suggests that its improvement cannot be explained solely by simple elevation-based correction. It also exhibits weaker systematic dependence on orography than the GSM–KF.

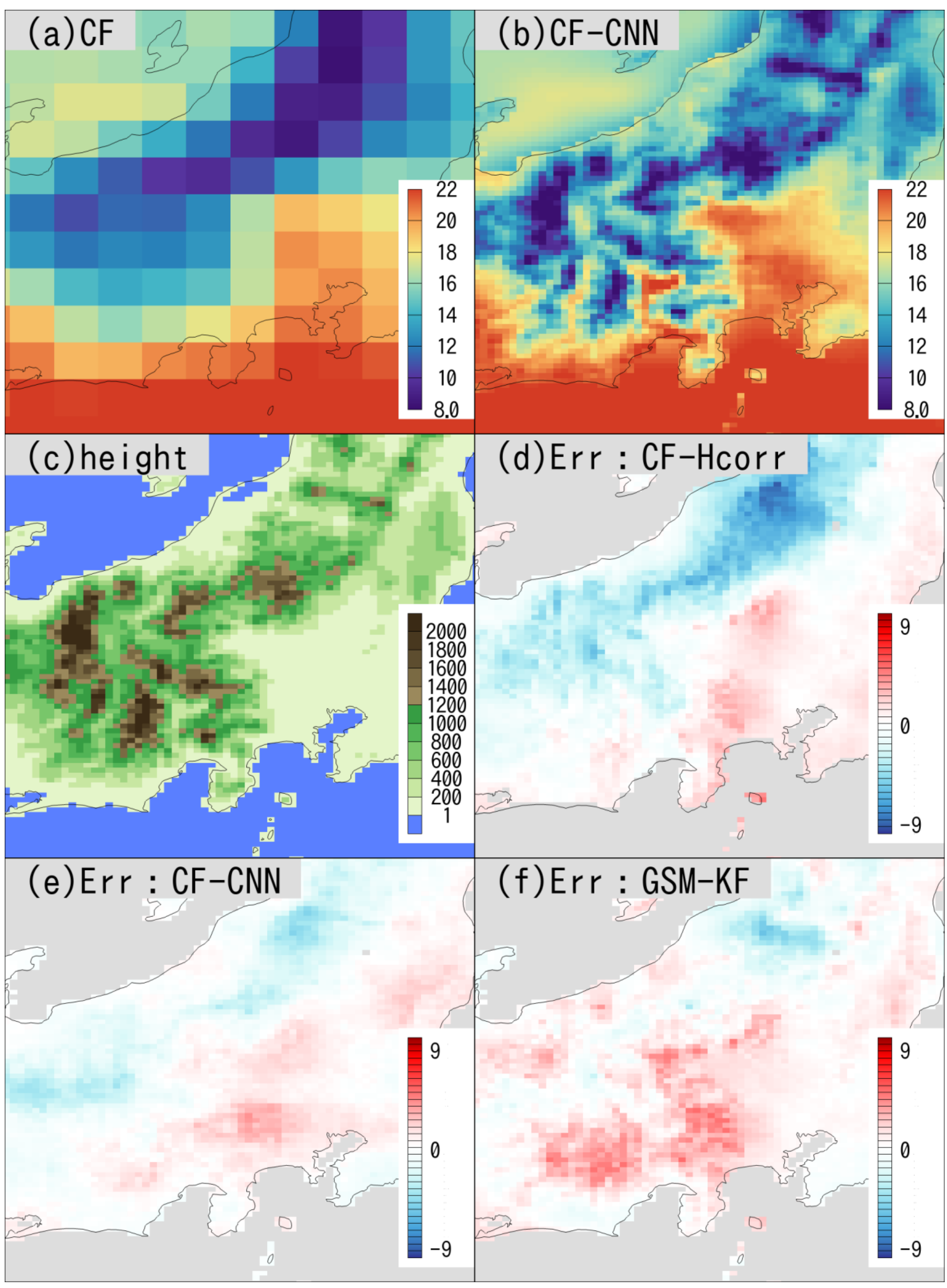

Fig. 10. Comparison of surface temperature (°C) and forecast error (°C) at 1200 UTC (2100 LST) on 13 June 2022. (a) CF and (b) CF–CNN initialized at 0000 UTC on 10 June

2022, which corresponds to an 84-h lead time. (c) Topography (m). (d) CF error after elevation correction (CF–Hcorr), (e) CF–CNN forecast error, and (f) GSM–KF forecast error.

2) NEAR-FREEZING TEMPERATURE FORECAST FOR SNOWFALL PREDICTION

The southern coast of eastern Japan is frequently affected during winter by low-pressure systems that bring heavy snowfall; these are referred to as “South-Coast Cyclones” (Araki 2019). In the Tokyo metropolitan area, such events disrupt transportation and societal activities; however, they are difficult to forecast because they require reliable predictions of precipitation and near-surface temperature. On the afternoon of 13 February 2022, the Ministry of Land, Infrastructure, Transport and Tourism issued an emergency warning for heavy snowfall (Ministry of Land, Infrastructure, Transport and Tourism 2022). However, the expected transition from rain to snow was delayed by a slower-than-expected nighttime temperature drop, and the heavy snowfall did not occur. Even slight deviations in surface temperature forecasts near the freezing point (0°C) can alter the snow-to-liquid ratio and precipitation type (e.g., snow or rain), leading to substantial differences in snowfall accumulation (Jennings and Molotch 2019; Furuichi and Matsuzawa 2009). As highlighted in this case, temperature forecast errors can cause substantial overestimations of snowfall events and their associated societal impacts.

The near-surface air-temperature threshold separating rain and snow generally ranges from approximately −1°C to 2.5°C, depending on regional and environmental factors such as humidity and elevation (Ye et al. 2013; Jennings et al. 2018). Because Japan has a humid midlatitude maritime climate, a relatively warm threshold of 2.5°C is adopted in this case study as the criterion to distinguish snow from rain. Figure 11a shows the GEPS forecast of the surface temperature at a lead time of 84 h. Both the CF and GEPS–EM, the latter representing the average of CF and 50 PF members, exhibit a 2.5°C isotherm bisecting the wide plain shown in Fig. 10c, suggesting a potential for snowfall over the lowlands. The GSM forecast with a higher horizontal resolution of 20 km (not shown), which employs the same physical parameterization schemes as the CF, shows a similar pattern. However, the CNN-corrected forecasts (CF–CNN and PF–CNNs) and their ensemble mean (CNN–EM) show differences (Fig. 11b). The CF–CNN predicts temperatures below 2.5°C over roughly half of the plain, whereas the CNN–EM predicts temperatures exceeding 2.5°C over almost the entire plain, consistent with the EST (Fig. 11c).

The error fields in Figs. 11d–f illustrate the advantage of the CNN-based forecasts in this case. Both single-member correction (CF–CNN; Fig. 11d) and ensemble-mean correction (CNN–EM, Fig. 11e) show smaller errors over the plain than the operational GSM–KF (Fig. 11f). A key difference, however, lies in the placement of the 2.5°C isotherm (red contours). The CF–CNN still does not align the 2.5°C isotherm well along the complex terrain boundary. Figure 11b also shows that some individually corrected PF–CNNs place the 2.5°C isotherm less accurately. However, when these corrected members are averaged, the resulting CNN–EM places the 2.5°C isotherm along the transition from the mountains to the plain, which is in closer agreement with the EST.

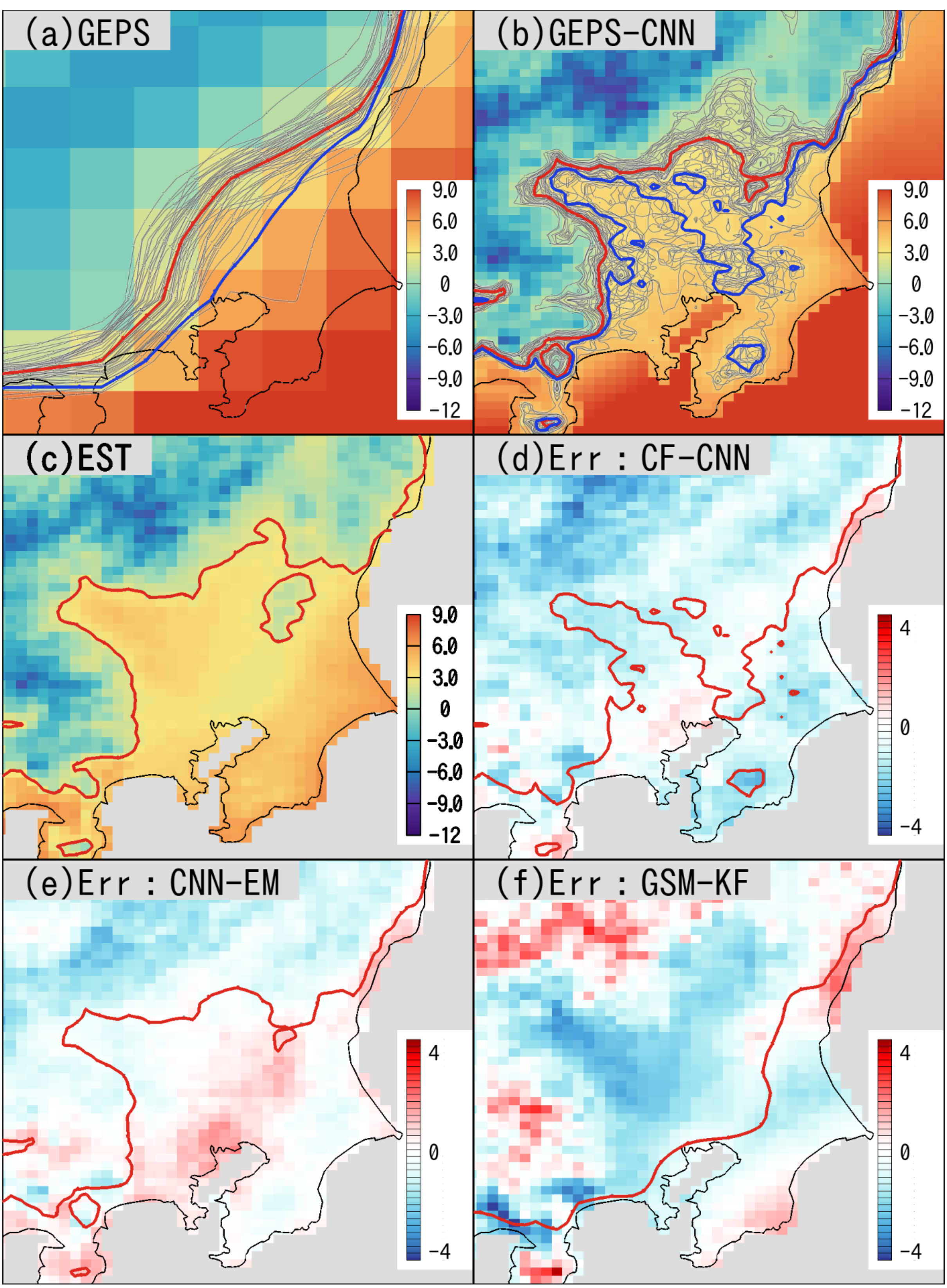

Fig. 11. Comparison of surface temperature (°C) and forecast error (°C) at 1200 UTC (2100 LST) on 13 February 2022. In all panels, contours indicate the 2.5°C isotherm. (a) Colors show the GEPS–EM, and the blue, gray, and red contours correspond to CF, PF, and GEPS–EM, respectively. (b) Colors represent the CNN–EM, and the blue, gray, and red contours correspond to CF–CNN, PF–CNN, and CNN–EM, respectively. (c) EST. Forecast errors for (d) CF–CNN, (e) CNN–EM, and (f) GSM–KF. All forecasts in (a) and (b) are initialized at 0000 UTC on 10 February 2022 with a lead time of 84 h. Colors denote forecast errors, and red contours indicate the 2.5°C isotherm.

## 6. Discussion

*a. Smoothing or Improving?*

To address the second part of Q3, we examine how the error correction of CNN-based post-processing differs from the smoothing effects of ensemble averaging. Following BG25, we decompose forecast performance into meaningful information (FI) and random error (NE) to provide insights beyond conventional skill metrics. Figure 12a shows an FI–NE–ACC diagram similar to that in Fig. 3 of BG25. BG25 plots the information and noise metrics defined by FTZP24, whereas FI and NE defined by BG25 are used herein.

Deterministic forecasts (CF, GSM, and MSM) exhibit a gradual decrease in FI and an increase in NE with increasing forecast lead time while maintaining their activity (distance from the origin). In contrast, the EM forecasts (GEPS–EM and CNN–EM) show decreasing FI, NE, and activity with increasing forecast lead time, eventually approaching the origin. This indicates that the differences in the characteristics of deterministic and EM forecasts (Fig. 10a) are mostly consistent with those reported in BG25. However, the CNN–EM exhibits a higher ACC compared with the CF–CNN, consistent with that in BG25; in contrast, the ACC of the GEPS–EM remains nearly identical to that of the CF.

An analysis of FI and NE shows that FI increases as the model resolution improves from 40 km for the CF to 20 km for the GSM and 5 km for the MSM. The ordering of NE among the forecasts is consistent with that of the RMSE shown in Fig. 5, indicating that NE appropriately represents the overall magnitude of forecast errors.

Figure 12b presents a comparison of the behaviors of EM (GEPS–EM and CNN–EM), KF-based correction (GSM–KF), and CNN-based post-processing (GSM–CNN, CF–CNN, and PF–CNN); in the figure, the axes are rescaled and the forecast lead time is restricted to 84 h. The FI of the GEPS–EM is comparable to the mean FI of the PFs, and both FI and NE are smaller than those of the CF (the same relation holds among the CNN–EM, PF–CNNs, and CF–CNN). This indicates that FI and NE are reduced via ensemble averaging. Furthermore, applying the KF or CNN to the GSM reduces NE. Even compared with the reduction achieved by ensemble averaging, the reductions produced by the KF and CNN are larger, with the CNN yielding the largest decrease. This is consistent with the RMSE results shown in Fig. 5a. The mean FI of the GSM–KF and GSM–CNN is slightly smaller than that of the GSM, but this decrease remains within the 95% confidence interval of the GSM. In contrast, the CF–CNN increases the FI of the CF. Consequently, these post-processed 5-km forecasts (GSM–KF, GSM–CNN, and CF–CNN) converge to a similar FI level. Meanwhile, although the CNN–EM is also a 5-km forecast, its FI is smaller than those of the GSM–KF, GSM–CNN, and CF–CNN. These results suggest that CNN-based post-processing and ensemble averaging reduce forecast errors in qualitatively different ways. Among the approaches examined here, the CNN can be regarded as the post-processing method that most strongly reduces NE while maintaining and, in some cases, increasing FI.

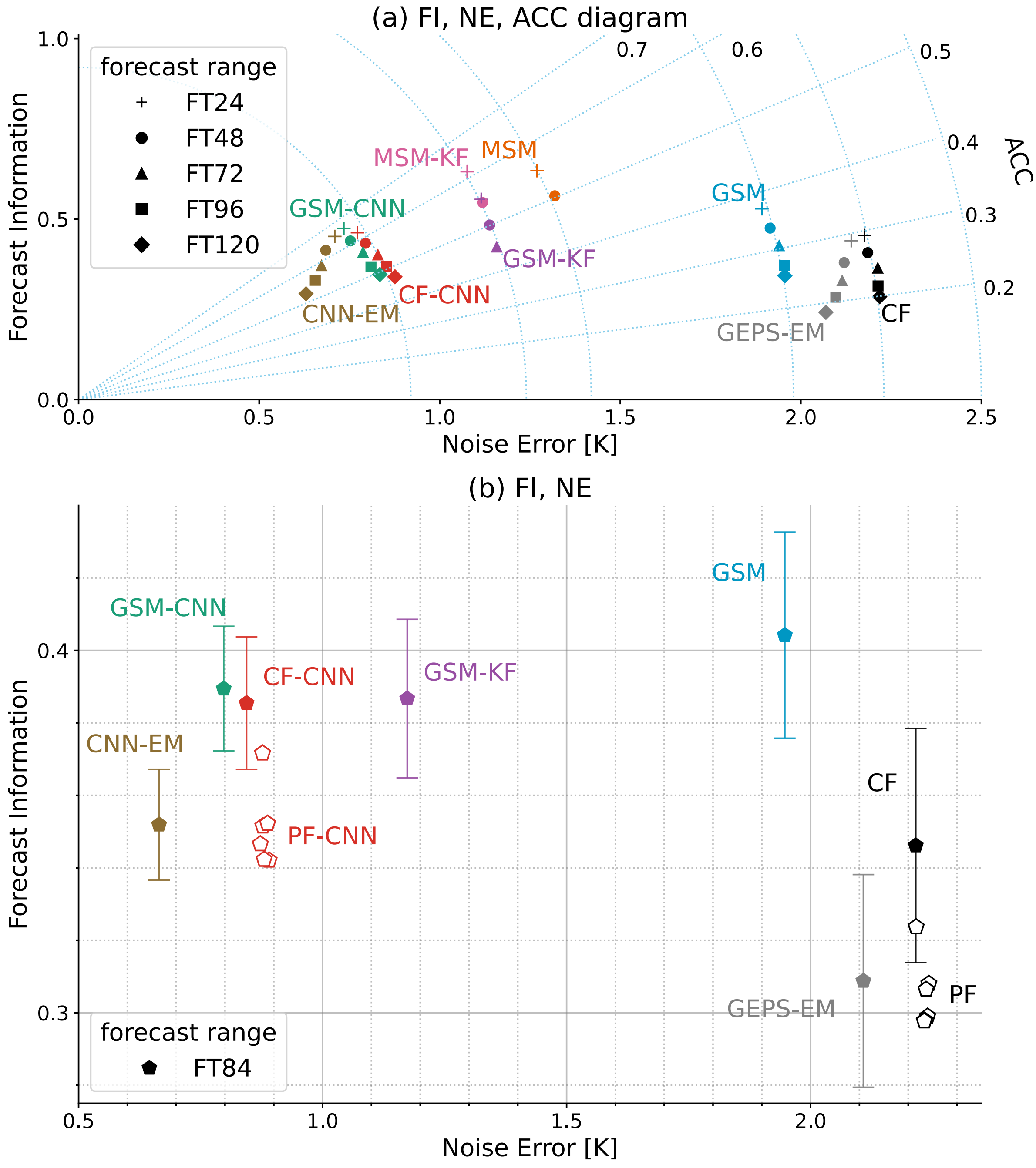

Fig. 12. (a) FI–NE–ACC diagram and (b) FI–NE plots at representative forecast lead times. The definitions of FI and NE follow those reported by Bonavita and Geer (2025).

### *b. Limitations: Prediction of Extreme Temperatures*

Ensemble averaging inherently smooths forecast fields, which suppresses extreme temperatures and limits the ability to forecast extreme events such as heatwaves. In the previous section, we demonstrate that CNN-based post-processing reduces forecast errors without information degradation by smoothing. Nevertheless, limitations remain for pronounced extremes. In this section, we present an example of an extreme heat event for which CNN-based post-processing still exhibits limitations that are not due to artificial smoothing but rather different factors.

An extraordinary heatwave occurred on 30 June 2022, and the observed surface temperatures exceeded 39°C. Figure 13 compares the surface temperature forecasts for this extreme event, initialized at 0000 UTC on 27 June 2022 and verified at a lead time of 78 h, corresponding to the maximum forecast range of the operational MSM and MSM–KF. Panels (a), (b), and (c) show the GEPS forecast, GEPS–CNN forecast, and EST, respectively. The low-resolution 40-km GEPS forecast (Fig. 13a), with shading for GEPS–EM and contoured for 34°C isotherm, fails to capture the extremely high temperatures observed over the plains (Fig. 13c). The 34°C contour appears only in a few PFs (gray) but not in the GEPS–EM (red). After CNN correction (Fig. 13b), both CNN–EM (red) and the CNN-corrected 36th PF (PF36–CNN; blue) show an extended high-temperature region compared with the original PF36. This suggests that the CNN-based correction improves the spatial extent of the warm region; however, the CNN–EM underestimates the maximum intensity.

These forecasts are compared with post-processed deterministic forecasts derived from the 20-km GSM and 5-km MSM. GSM forecasts are downscaled to 5-km resolution using either KF- or CNN-based post-processing. The 5-km GSM–KF forecast (not shown) more easily represents localized extremes by first applying pointwise bias correction with the KF and then interpolating to the grid. The GSM–CNN (not shown) also shows high-temperature areas more clearly than the GEPS (Fig. 13b), although the predicted temperatures remain lower than the observations. These results indicate that post-processing with either the KF or CNN cannot fully compensate for the underrepresentation of high temperatures in the GSM input fields. In contrast, the MSM–KF forecast (Fig. 13d), derived from a regional NWP model (MSM) independent of the GSM, more closely reproduces the observed warm-area extent.

For this case, the GEPS–CNN reduces the ensemble spread slightly from 0.93 to 0.84 K (−0.09 K), whereas the CRPS improves substantially from 3.19 to 2.62 K (−0.57 K). This indicates that the improvement in the overall distribution far outweighs the modest reduction in the ensemble spread. To further diagnose these distributional differences, we use quantile–quantile (Q–Q) plots. For each forecast field $F$ and observation $O$, the quantiles $q_p(F)$ and $q_p(O)$ are computed for probability levels $p \in [0.01, 0.99]$ at an interval of 0.01. These forecast quantiles are plotted against the corresponding observed quantiles.

The Q–Q plot of temperature (Fig. 14) confirms these differences. For the GEPS–EM, the curve lies below the diagonal for most of the distribution, which indicates a general negative bias. This deficiency is particularly pronounced in the warm-side upper tail (the cross at $p = 0.99$), which signifies a notable underestimation of the extreme heat. In contrast, the Q–Q plot for the CNN–EM shows that the upper end on the warm side and the overall slope of the curve are closer to the diagonal, indicating improved reproduction of temperature distribution, particularly in the upper-tail. However, the curve remains at a roughly constant distance below the diagonal, indicating that negative biases remain. The MSM–KF, which employs the higher-resolution MSM rather than the GEPS as the input, lies closest to the diagonal in both its overall slope and end points, and it reproduces the observed distribution well over a wide range of quantiles. Therefore, the failure to fully reproduce the most extreme heat is not a result of artificial smoothing but is instead due to limitations of the CNN itself and the limited information contained in the low-resolution GEPS inputs.

Although CNN correction shows strong potential for enhancing spatial consistency and generalization and has even been successfully applied to post-processing forecasts of extreme high temperatures (e.g., Cho et al. 2022), this case study suggests that the CNN architecture should be further improved and ensemble information should be more effectively used for more reliable predictions of pronounced temperature extremes.

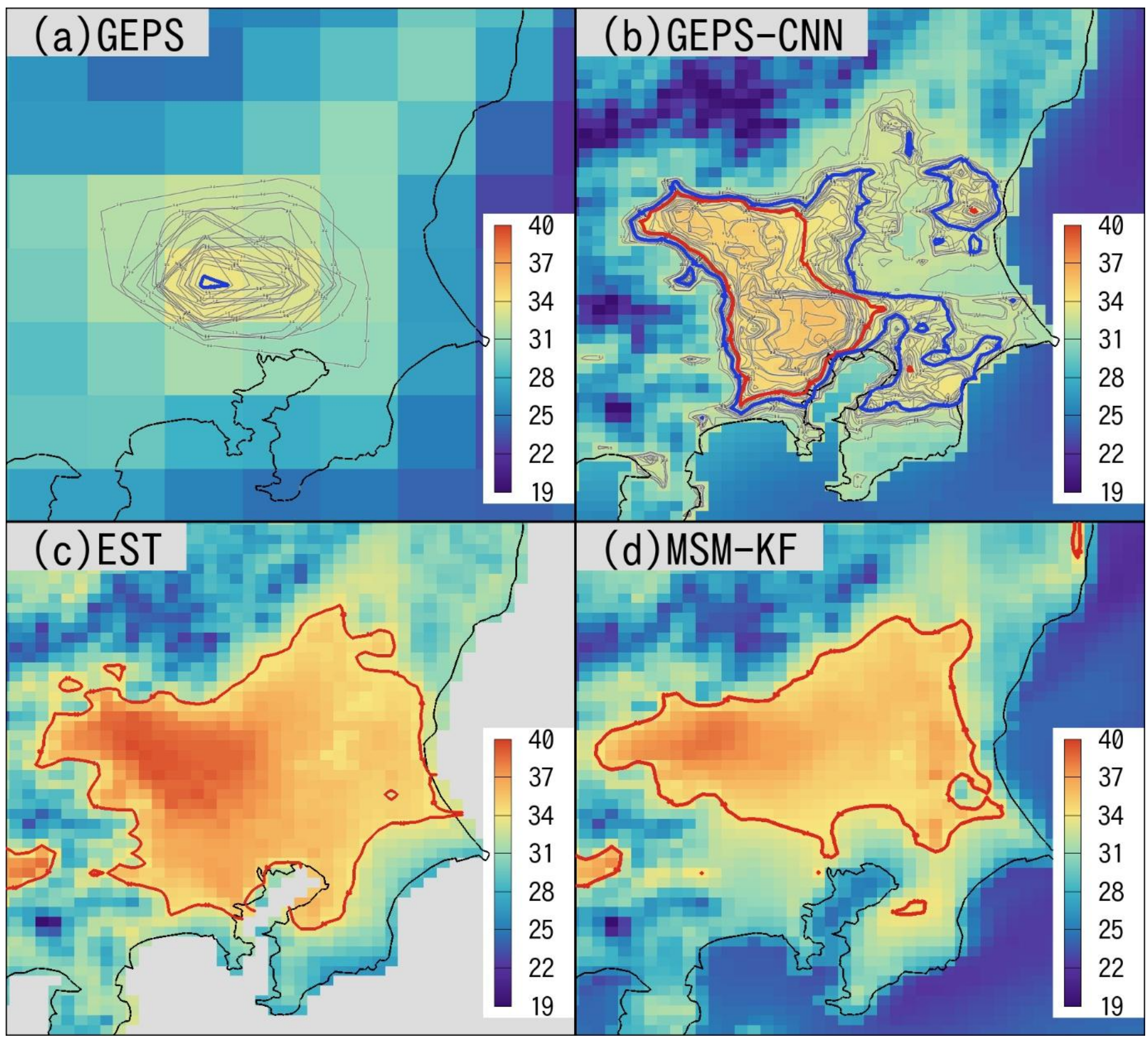

Fig. 13. Surface temperatures (°C) for the extreme heat event at 0600 UTC (1500 LST) on 30 June 2022 (forecasts are initialized at 0000 UTC 27 June 2022 with a 78-h lead time). Color shades represent the (a) GEPS–EM, (b) CNN–EM, (c) EST, and (d) MSM–KF. The contours indicate the 34°C isotherm. In (a), gray, red, and blue contours correspond to the PFs, GEPS–EM, and 36th perturbed member (PF36), respectively; in (b) these contours correspond to the PF–CNNs, CNN–EM, and PF36–CNN, respectively.

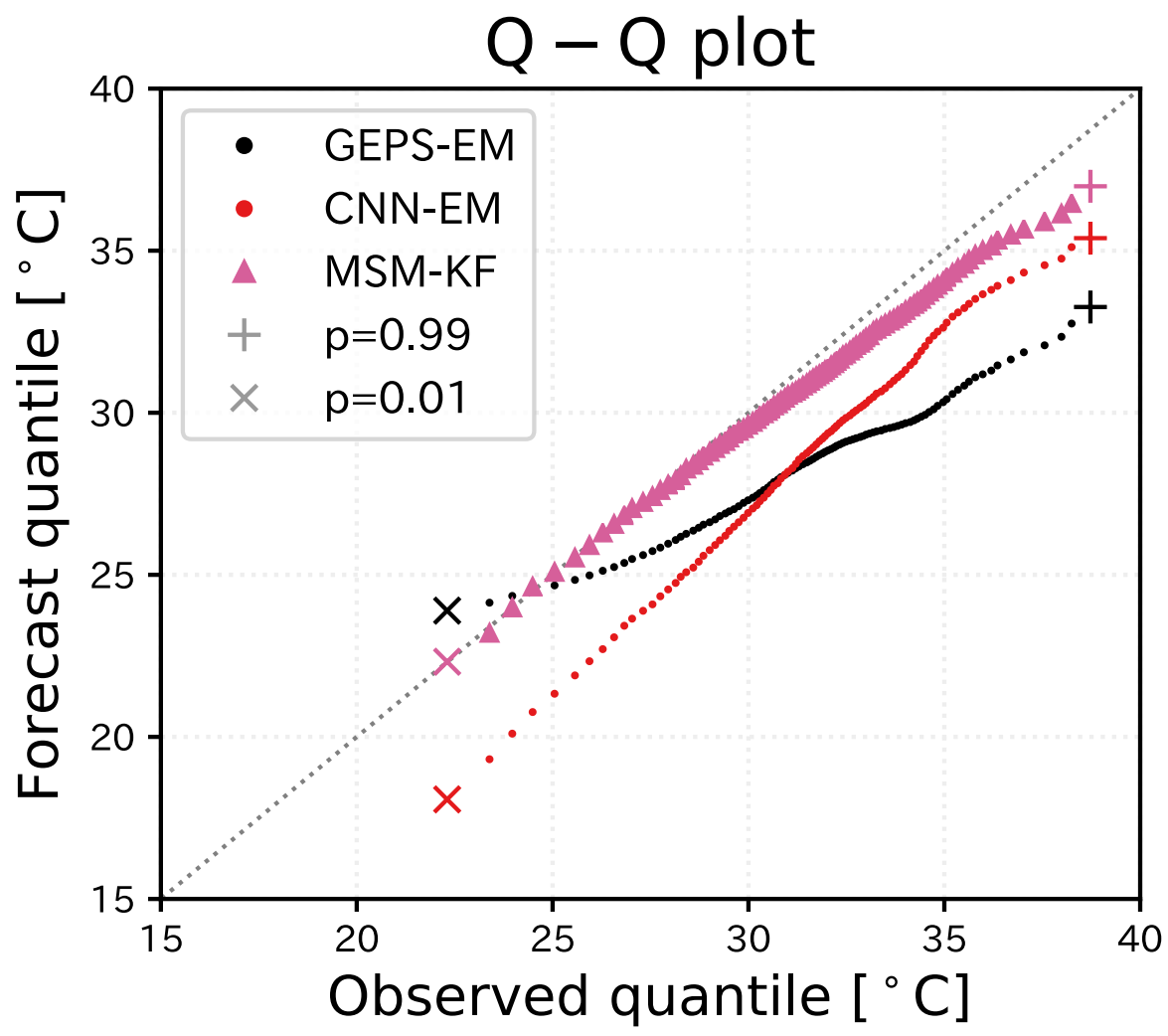

Fig. 14. Q–Q plot of temperatures for the 78-h forecast (initialized at 0000 UTC 27 June 2022). Dots show quantile pairs between observations and forecasts for the GEPS–EM (black) and the CNN–EM (red) and triangles for the MSM–KF (pink). The 1:1 line (gray dotted) indicates perfect agreement. Symbols "+" and "×" indicate the upper ($p = 0.99$) and lower ($p = 0.01$) distribution tails, respectively.

## 7. Conclusion

To enhance medium-range surface temperature predictions, a post-processing method that integrates CNN-based bias correction and downscaling with ensemble forecasting is proposed herein. By operating on low-resolution global ensemble model outputs and producing high-resolution forecasts, the CNN overcomes the spatial-resolution limitation of existing ensembles and enables 5-km-class medium-range forecasts under limited computational resources; this is an important advantage for operational systems that rely on low-resolution models to extend their forecast range. Based on these results, the three questions posed in the Introduction can be answered as follows.

A1. (How effectively can a CNN reduce systematic and random errors?): Section 5a shows that CNN-based post-processing achieved substantial improvements in deterministic and ensemble forecasts. When applied to deterministic forecasts from the JMA global and regional models (GSM and MSM), the CNN substantially reduced the RMSE and ME at all lead times. The CF–CNN improved the lead time–averaged RMSE by 1.2 K (46%) and reduced the corresponding ME by 0.18 K, outperforming all operational deterministic NWP models and their KF-based methods (GSM–KF and MSM–KF). When combined with ensemble averaging, member-wise CNN correction synergistically reduced both systematic and random forecast errors.

A2. (How does training member selection influence performance?): Section 5b demonstrates that CNNs trained individually using the CF, PF, and GEPS–EM showed similar performances when applied to all ensemble members. These CNNs reduced the lead time–averaged RMSE of both the CF and PFs, and the resulting ensembles exhibited similar improvements in the SSRs. No statistically significant differences were observed among the three training strategies. These results indicate that training the CNN on a single representative member such as the CF is sufficient. This approach considerably simplifies the design and operational implementation of CNN-based post-processing for ensemble prediction systems.

A3. (How does CNN improve ensemble performance and differ from smoothing effects?): Section 5c reveals that the CNN improved the ensemble reliability, as shown by the decreased CRPS, the SSR approaching 1.0, and flatter conditional rank histograms after CNN-based post-processing. Furthermore, analysis using FI and NE indicates that CNN-based post-processing and ensemble averaging improved the forecasts in qualitatively different ways. Whereas ensemble averaging reduced both FI and NE through smoothing, CNN-based post-processing mainly reduced NE while maintaining an FI value comparable to those of other deterministic 5-km forecasts and, in some cases, increasing it. These forecast improvements therefore resulted from genuine error reduction rather than artificial skill gains via excessive smoothing. This clarifies the relation between CNN-based post-processing and the smoothing effects of ensemble averaging.

Case studies highlighted the practical benefits of the proposed method. Over complex terrain in central Japan, the CNN produced more spatially detailed temperature patterns consistent with the 5-km topography and reduced biases beyond those removed by a simple elevation-based correction. In the case of a near-freezing “South-Coast Cyclone”, the CNN provided a more realistic representation of the rain–snow transition at a lead time of 84 h with the ensemble-mean correction showing closer agreement with the EST than the single-member correction. These examples suggest that CNN-based bias correction and downscaling can enhance the spatial detail and temporal consistency in medium-range forecasts in situations where small temperature errors have large societal impacts.

Nonetheless, the proposed method has limitations in predicting extreme events. In an extraordinary heatwave case, the CNN expanded the spatial extent of high temperatures and improved the upper-tail distribution compared with the original ensemble mean but still underestimated the maximum intensity. This suggests that the performance of CNN-corrected forecasts in such extremes is constrained by the limited predictability of the low-resolution NWP inputs. Although our experiments were limited to Japan, the proposed method is not inherently Japan-specific and can be transferred to other regions given suitable reference analyses, with retraining or domain adaptation as needed. In future work, incorporating probabilistic forecasting frameworks built on CNN-corrected ensemble outputs may further enhance the reliability of uncertainty quantifications (Rasp and Lerch 2018; Dueben and Bauer 2018; Sønderby et al. 2020). In addition, exploring alternative deep learning architectures such as vision transformers (Dosovitskiy et al. 2021) to better capture long-range spatial dependencies than CNNs may improve forecasting in complex or highly nonlinear scenarios (Pathak et al. 2022; Lam et al. 2023).

*Acknowledgments.*

We sincerely thank Mr. S. Ito and Mr. Y. Shirayama of the Numerical Prediction Development Center at the Japan Meteorological Agency (JMA) and Dr. T. Sekiyama and Dr. D. Hotta of the Meteorological Research Institute (MRI) at JMA for their valuable discussions. This work was supported by the Japanese Society for the Promotion of Sciences KAKENHI (Grants 24H00278, 25H00784, and 25K01483) and by a Fugaku General Access Project (Project ID: hp250210).

*Data Availability Statement.*

The source code of the CNN models employed in this study is available upon request, subject to collaborative research agreement with the MRI. Numerical weather-prediction model output datasets of the JMA are operationally provided via the Japan Meteorological Business Support Center (JMBSC; http://www.jmbsc.or.jp/en/index-e.html). Datasets not available through JMBSC may be obtained through collaborative research agreement with the MRI and/or the JMA.

APPENDIX

## Appendix A

Table A1. Functions and parameters of the CNN shown in Fig. 1 (based on Table 1 from Inoue et al. 2024).

| Unit | Function | Parameters |
|---|---|---|
| Conv1 | Conv2d | kernel_size = 5, stride = 1, padding = 2, number of channels: 7 to 32 |
| | MaxPool2d | kernel_size = 2, stride = 2 |
| | BatchNorm2d | number of channels: 32 |
| | ReLU | |
| Conv2 | Conv2d | kernel_size = 5, stride = 1, padding = 2, number of channels: 32 to 64 |
| | MaxPool2d | kernel_size = 2, stride = 2 |
| | BatchNorm2d | number of channels: 64 |
| | ReLU | |
| FC1 | Linear | number of units: 65536 to 4096 |
| | BatchNorm1d | number of units: 4096 |
| | ReLU | |
| FC2 | Linear | number of units: 4096 to 65536 |
| | BatchNorm1d | number of units: 65536 |
| | ReLU | |
| ConvT1 | ConvTranspose2d | kernel_size = 2, stride = 2, padding = 0, number of channels: 64 to 32 |
| | BatchNorm2d | number of channels: 32 |
| | ReLU | |

| ConvT2 | ConvTranspose2d | kernel_size = 2, stride = 2, padding = 0, number of channels: 32 to 1 |
|---|---|---|
| | BatchNorm2d | number of channels: 1 |
| | Sigmoid | |